\documentstyle[12pt,a4wide]{article}
\begin{document}

\newcommand{\ds}{\displaystyle}

\newcommand{\zz}{\cal Z}
\newcommand{\uz}{ {\bf z} }
\newcommand{\UZ}{ {\bf Z} }
\newcommand{\pr}{\prime}
\newcommand{\rr}{{\cal R}}

\newcommand{\th}{\theta}
\newcommand{\tb}{\bar{\theta}}
\newcommand{\TH}{\Theta}
\newcommand{\TB}{\bar{\Theta}}
\newcommand{\la}{\lambda}
\newcommand{\LA}{\Lambda}
\newcommand{\LB}{\bar{\LA}}
\newcommand{\OM}{\Omega}
\newcommand{\DE}{\Delta}
\newcommand{\bt}{\bar{\tau}}
\newcommand{\zb}{\bar z}
\newcommand{\pa}{\partial}
\newcommand{\dab}{\bar D}
\newcommand{\ZB}{\bar Z}
\newcommand{\pab}{ \bar{\partial} }
\newcommand{\HT}{ {H_{\th}}^z }
\newcommand{\HB}{ {H_{\tb}}^z }
\newcommand{\HO}{ H_{\th} ^{\ \th} }
\newcommand{\HZ}{ H_{\zb} ^{\ \th} }
\newcommand{\HZB}{ H_{\zb} ^{\ z} }
\newcommand{\HOB}{ H_{\tb} ^{\ \th} }

\newcommand{\zt}{\tilde z}
\newcommand{\dt}{\tilde D}
\newcommand{\pabt}{\tilde{\pab}}

\newcommand{\thm}{\theta^-}
\newcommand{\tbm}{{\bar{\theta}}^-}

\newcommand{\lb}{\bar{\la}}
\newcommand{\kb}{\bar k}
\newcommand{\al}{\alpha}
\newcommand{\ab}{\bar{\al}}
\newcommand{\be}{\beta}
\newcommand{\bb}{\bar{\be}}
\newcommand{\ga}{\gamma}
\newcommand{\gb}{\bar{\ga}}
\newcommand{\de}{\delta}
\newcommand{\db}{ \bar{\de}}
\newcommand{\qb}{{\bar Q}}
\newcommand{\mb}{{\bar{\mu} }}

\newcommand{\nab}{\nabla}
\newcommand{\nabar}{\bar{\nabla}}

\newcommand{\pal}{{\rm pal}}

\newcommand{\vv}{{\cal V}}


\thispagestyle{empty}

\hfill  LYCEN-PUB97-30
\vskip 0.07truecm
\hfill MPI-PhT/97-36
\vskip 0.07truecm
\hfill  solv-int/9708???
\vskip 0.09truecm
\hfill  August 1997

\begin{center}
{\bf \Huge{$d=2, \, N=2$}}
\end{center}
\begin{center}
{\bf \Huge{Superconformally Covariant Operators}}
\end{center}
\begin{center}
{\bf \Huge{and}}
\end{center}
\begin{center}
{\bf \Huge{Super $W$-Algebras}}
\end{center}

\vskip 0.9truecm
\centerline{{\bf
Fran\c cois Gieres}$\, ^{a, b,\S}$ 
$\quad$ and $\quad$
{\bf St\'ephane Gourmelen}$\,^a$}
\bigskip
\bigskip
\centerline{$^a$ {\it Institut de Physique Nucl\'eaire de Lyon,
IN2P3/CNRS}}
\centerline{\it Universit\'e Claude Bernard}
\centerline{\it 43, boulevard du 11 novembre 1918}
\centerline{\it F - 69622 - Villeurbanne Cedex}
\ \\
\centerline{$^b$ {\it Max-Planck-Institut f\"ur Physik}}
\centerline{ {\it -- Werner Heisenberg Institut --}}
\centerline{ {\it F\"ohringer Ring 6}}
\centerline{ {\it 80805 Munich (Fed. Rep. Germany)}}
\vskip 0.9truecm

\nopagebreak
\begin{abstract}

We construct and classify superconformally covariant
differential operators defined on $N=2$ super Riemann surfaces.
By contrast to the $N=1$ theory, these operators
give rise to partial rather than ordinary differential equations
which leads to novel features for their matrix representation.
The latter is applied to the derivation of $N=2$ super
$W$-algebras in terms of $N=2$ superfields.

\end{abstract}
\bigskip

\nopagebreak
\begin{flushleft}
\rule{2 in}{0.03cm} \\

{\footnotesize \ ${}^{\S}$
Alexander von Humboldt Fellow on leave of absence from the 
Universit\'e Claude Bernard.
} \\	 
\end
{flushleft}


\section{Introduction}

From reference \cite{cco}, we recall
that the {\em Bol operator} $L_n$ (acting on a compact Riemann surface
with local coordinates $z$)
is the conformally covariant
version of the differential operator
$\pa^n \equiv ({\pa \over \pa z}) ^n$. The simplest examples
are given by
\begin{eqnarray}
L_0 & = & {\rm 1}
\nonumber  \\
L_1 & = & \pa
\nonumber  \\
L_2 & = & \pa^2 \ + \ \frac{1}{2} \, R
\nonumber  \\
L_3 & = & \pa^3 \ + \ 2 \, R \, \pa \ + \ (\pa R)
\label{81}
\\
L_4 & = & \pa^4 \ + \ 5 \, R \, \pa^2 \ + \ 5\, (\pa R)\, \pa \ + \
\frac{3}{2} \, \left[ (\pa^2 R) \, + \, \frac{3}{2} \, R^2  \right]
\ \ ,
\nonumber
\end{eqnarray}
where $R = R_{zz} (z)$
denotes a projective connection, i.e. a locally
holomorphic field transforming with the Schwarzian derivative
under a conformal change
of coordinates $z \to z^{\prime} (z)$:
\begin{equation}
\label{int}
R^{\prime} (z^{\prime} ) = (\pa z^{\prime} )^{-2}
\left[ R (z) - S(z^{\prime} , z) \right]
\quad {\rm where} \quad
S(z^{\prime} , z) =
\pa^2  {\rm ln} \, \pa z^{\prime}
-{1\over 2} ( \pa \, {\rm ln} \, \pa z^{\prime}  )^2
\ .
\end{equation}
The most general conformally covariant operator of order $n$
whose highest coefficient is normalized to one 
can be written as a sum of $L_n$ and some lower order operators
$M^{(n)}_{w_k}$
depending linearly on some conformal fields $w_k$ (with $k = 3,..,n$),
e.g.
\begin{eqnarray}
\label{mwb}
M^{(n)}_{w_n} & = & w_n
\qquad , \qquad
M^{(n)}_{w_{n-1}} \; = \; w_{n-1} \pa + {1\over 2} (\pa w_{n-1} )
\\
M^{(n)}_{w_{n-2}} & = & w_{n-2} \left[ \pa^2 -  {1-n \over 2} R \right]
+ (\pa w_{n-2} ) \pa
+ {n-1 \over 2(2n-3)} \left[\pa^2  - (n-2) R \right]  w_{n-2}
\ \ .
\nonumber
\end{eqnarray}
The operators $L_n$ and
$M^{(n)}_{w_k}$ can both be obtained from M\"obius covariant operators
and they admit a matrix representation which is related to the
principal embedding of $sl(2)$ into $sl(n)$ \cite{dfiz}. All of these
covariant operators admit numerous applications to conformal and
integrable models, in particular to the theory of $W$-algebras.
Their $N=1$ supersymmetric generalization has been worked
out in references \cite{cco, gt1} (see also \cite{hua})
and the present paper is devoted
to their $N=2$ supersymmetric generalization.

There is a new feature in $N=2$ superspace geometry which makes this
theory considerably richer and more complicated:
the ``square root" of the translation generator $\pa$
is not given by a single odd operator $D_1$ as in $N=1$ supersymmetry
($D_1^2 = \pa$), but it involves two odd operators
($ \{ D , \dab \} = \pa$).
Therefore, one has to deal with partial differential equations
(involving $D$ and $\dab$) rather than ordinary differential
equations (only involving $D_1$).
Another aspect of the algebra
$ \{ D , \dab \} = \pa$ consists of the fact that it introduces a
$U(1)$ symmetry into the theory: after projection from the
super Riemann surface to the underlying ordinary Riemann
surface, one thereby recovers $U(1)$-transformations
in addition to the familiar conformal transformations. Henceforth,
the operators (\ref{81})(\ref{mwb}) acting on $U(1)$-neutral fields
are to be generalized to conformally covariant operators
acting on $U(1)$-charged fields.
The latter as well as the original operators
(\ref{81})(\ref{mwb}) arise from different classes of $N=2$ superdifferential 
operators.
In particular, one is led to introduce operators which relate
the chiral and anti-chiral subspaces
of superconformal fields.
It is only for these so-called `sandwich operators' that we will
be able to give a matrix representation. The latter provides
a gauge connection field
 with values in the Lie superalgebra
$sl(n+1 |n)$ and the associated
$N=2$ super $W$-algebra \cite{lu} can be constructed by imposing
zero curvature conditions on this connection.

To keep supersymmetry manifest,
all of our considerations will be carried out in superspace while the
projection to ordinary space will be indicated in each case.
The passage from the $N=2$ to the $N=1$ superfield formalism
will also be discussed so as to make contact with previous work
which mostly uses the latter description.
Applications of our results will be mentionned in the text and in the
concluding remarks.

\section{Geometric framework}

In this section, we
recall some definitions
(following \cite{dgg} and references therein)
and we introduce some tools which are needed
in the sequel.

\subsection{Notation and basic relations}

We will work on
a compact $N=2$ super Riemann surface (SRS)
locally parametrized by coordinates
\begin{equation}
\label{coo}
( \uz ; \bar{\uz} ) \equiv (z, \th, \tb ;
\zb , \thm, \tbm ) \equiv
(x^{++} , \th^+ , \tb^+ ;
x^{--} , \thm, \tbm )
\ \ ,
\end{equation}
with $z, \zb$ even and $\th, \tb, \thm, \tbm$ odd.
The variables are complex and related by complex conjugation
(denoted by $\ast$):
\[
z^{\ast} = \zb
\quad , \quad
(\th^+ )^{\ast} = \th^-
\quad , \quad
(\tb^+ )^{\ast} = \tb^-
\ \ .
\]
As indicated in (\ref{coo}), we will
omit the plus-indices of $\th^+$ and $\tb^+$ to simplify
the notation.

The canonical basis of the
tangent space is defined by
$( \pa , \,
D  , \,
\dab ; \,
\pab , \,
D_-  , \,
\dab _- )$  with
\begin{eqnarray}
\label{1}
\pa  & =& \frac{\pa}{\pa z}
\quad , \quad
D \ = \ \frac{\pa}{\pa \th}  +  \frac{1}{2} \, \tb \pa
\qquad \ , \quad  \
\dab \ = \ \frac{\pa}{\pa \tb}  +  \frac{1}{2} \, \th  \pa
\quad {\rm and \ c.c.}
\ .
\end{eqnarray}
The graded Lie brackets between these vector fields are given by
\begin{equation}
\label{2}
\{  D ,  \dab \} =  \pa
\qquad \qquad
{\rm and \ c.c.}
\ \  ,
\end{equation}
all others brackets being zero, in particular,
\begin{equation}
\label{3}
D^2   =  0  = \dab^2
\qquad \qquad
{\rm and \ c.c.}
\ \ .
\end{equation}
Note that this set of equations implies 
$[D,\dab ]^2 = \pa^2$ and c.c. \ .

\subsection{Superconformal transformations}

By definition of the SRS,
any two sets of local coordinates, say  $\uz$
and $\uz^{\pr}$,
are related by a
{\em superconformal transformation}, i.e.
a smooth mapping satisfying the three conditions

\begin{eqnarray}
{\rm (i)} & \quad &
{\uz}^{\prime}  =  {\uz} ^{\prime} ( {\uz} )
\quad \Longleftrightarrow \quad
D_- {\uz}^{\prime}  = 0 =
\dab_- {\uz}^{\prime}
\nonumber
\\
\label{3i}
{\rm (ii)} & \quad &
D \tb^{\prime} \ = \ 0 \ =
\dab \th^{\prime}
\\
{\rm (iii)} & \quad &
D z^{\prime}  =
\frac{1}{2}
\tb ^{\prime} \, D \th ^{\prime}
\quad \qquad , \quad \
\dab z^{\prime} \ = \ \frac{1}{2}
\th ^{\prime} \, \dab \tb ^{\prime}
\nonumber
\ \ .
\nonumber
\end{eqnarray}
These relations imply
that $D$ and $\dab$ separately transform
into themselves,
\begin{eqnarray}
D ^{\prime} & = & {\rm e} ^w \ D
\nonumber
\\
\dab ^{\prime} & = & {\rm e} ^{\bar{w}} \ \dab
\label{7a}
\\
\pa ^{\prime} & = &
\{  D^{\prime}  ,  \dab ^{\prime}  \}  =
{\rm e} ^{w+\bar{w}}  \, [  \pa   +
( \dab w )  D  +
(D \bar{w} )  \dab   ]
\ \ ,
\nonumber
\end{eqnarray}
where
\begin{eqnarray}
\label{8a}
{\rm e}^{-w} & \equiv & D \th^{\prime}
\ \ \ \ \ , \ \ \ \ \
D w \ = \ 0
\\
{\rm e}^{-\bar{w}} & \equiv & \dab  \tb^{\prime}
\ \ \ \ \ , \ \ \ \ \
\dab  \bar{w} \ = \ 0
\nonumber
\end{eqnarray}
and
\begin{equation}
\label{w}
{\rm e}^{-w-\bar{w}} \, = \,
\pa z^{\prime}
\, + \, \frac{1}{2} \, \tb^{\prime} \, \pa \th^{\prime}
\, + \, \frac{1}{2} \, \th^{\prime} \, \pa \tb^{\prime}
\ \ .
\end{equation}
We note that eqs.(\ref{7a})(\ref{8a}) imply
$(D\th^{\pr})(D^{\pr} \th ) = 1 = (\dab \tb^{\pr} )(\dab^{\pr} \tb)$
and that analogous equations hold in the $\zb$-sector.

\subsection{$U(1)$ symmetry}

Equation (\ref{w}) determines the product of $D\th^{\prime}$
and $\dab \tb^{\prime}$, but not their quotient which is related
to the $U(1)$ automorphism group of the $N=2$ supersymmetry algebra.
To introduce $U(1)$-{\em transformations} parametrized by a 
superanalytic superfield $K$, we
decompose equation (\ref{w})
in a symmetric way,
\begin{eqnarray}
\label{dec}
D\th^{\prime} & = &
{\rm e}^{+K/2} \
\left( \pa z^{\prime}
+ \frac{1}{2} \, \tb^{\prime}  \pa \th^{\prime}
+ \frac{1}{2} \, \th^{\prime}  \pa \tb^{\prime} \right) ^{1/2}
\\
\dab\tb^{\prime} & = &
{\rm e}^{-K/2} \
\left( \pa z^{\prime}
+ \frac{1}{2} \, \tb^{\prime}  \pa \th^{\prime}
+ \frac{1}{2} \, \th^{\prime}  \pa \tb^{\prime} \right) ^{1/2}
\ \ ,
\nonumber
\end{eqnarray}
which implies $K=  {\rm ln} \,
\left( D\th^{\prime} /
\dab\tb^{\prime} \right) $.

\subsection{Superconformal fields}

In the following, we will consider
{\em superconformal fields} $C_{p,q} ( {\uz} , \bar{\uz} )$
transforming like
\begin{equation}
\label{tra}
C_{p,q}^{\prime} = {\rm e}^{pw+q \bar w} C_{p,q}
\qquad  ( \, p,q \in {\bf Z}/2
\ \ , \ \ p+q \in {\bf Z} \, )
\end{equation}
and having a Grassmann parity $(-)^{p+q}$.
The space of these fields is denoted by ${\cal F}_{p,q}$.
The pair $(p,q)$ will be called the {\em superconformal
weight} of $C_{p,q}$ while ${1 \over 2} (p+q)$
and ${1 \over 2} (p-q)$ will be referred to as
its {\em conformal} and $U(1)$ {\em weight} (or {\em charge}),
respectively -- see eq.(\ref{trac}) below for the origin of this terminology. 
Thus, $C_{p,q}$ is
{\em neutral} with respect to $U(1)$ if $p=q$.
Fields with another index structure (transforming also with
$w^*, \bar{w} ^*$) can be defined in
an analogous way.

From the transformation properties of derivatives and fields,
one concludes that the following  chirality conditions are
superconformally covariant:
\begin{equation}
D C_{p,0} = 0   \qquad {\rm or} \qquad \dab C_{0,p} = 0
\ \ .
\end{equation}
The space of these so-called {\em chiral} and  {\em anti-chiral}
fields will be denoted by ${\cal F}^c _{p,0}$ and
${\cal F}^a _{0,p}$, respectively.

Note that the quadratic constraints
$D\dab C_{0,-1} =0$ and
$\dab D C_{-1,0} =0$ are also covariant.

\subsection{Projection to component fields}

A generic $N=2$ superfield admits the $\th$-expansion
\begin{eqnarray}
\label{suf}
C ( {\bf z} , {\bf {\bar z}} ) & = &
c + \th \psi + \tb \bar{\eta} + \th \tb d \ + \
\th^- [ \gamma + \th e + \tb f + \th \tb \delta ]
\\
& & +
\tb^- [ \epsilon + \th g + \tb h
+ \th \tb \tau  ]
\ + \ \th^- \tb^-
[ m  + \th \varphi
+ \tb \bar{\chi} + \th \tb n  ] \ \ ,
\nonumber
\end{eqnarray}
where the component fields $c,..., n $ depend on
$z$ and $\zb$. Equivalently, these space-time fields can be introduced
by means of projection,
\[
C  \vert \; =c \quad , \quad
DC  \vert \; = \psi \quad , \quad
\dab C  \vert \; =\bar{\eta} \quad , \quad
[D, \dab ] C  \vert \; =-2d
\quad , \quad ...
\ \ ,
\]
where the bar denotes the projection onto the lowest component
of the corres\-ponding superfield.

Ordinary conformal and $U(1)$-transformations are related
to the expressions
\begin{eqnarray}
\label{spe}
(w + \bar w )  \vert & = & - \, {\rm ln} \, \pa z^{\pr}
\\
(w - \bar w )  \vert & = & - \, {\rm ln} \, {D\th^{\pr} \over
\dab \tb^{\pr} } \, \vert \ = \ -K  \vert \ \equiv \ -k
\ \ .
\nonumber
\end{eqnarray}
Here and in all following space-time equations, $z^{\pr}$ stands
for the lowest component of the superfield $z^{\pr} (\uz )$ and the
{\em fermionic contributions are never spelled out}.

If $C \equiv C_{p,q}$ belongs to ${\cal F}_{p,q}$, then its transformation law
(\ref{tra}) can be projected down to ordinary
space by taking into account eqs.(\ref{dec})(\ref{spe}):
\begin{equation}
\label{trac}
c_{p,q}^{\pr} =
(\pa z^{\pr})^{- {1 \over 2} (p+q)} \, {\rm e}^{-{1\over 2} (p-q)k} \,
c_{p,q}
\ \ .
\end{equation}
Obviously,
${1 \over 2} (p+q)$ (resp. ${1 \over 2} (p-q)$) is the conformal
(resp. $U(1)$) weight of the ordinary conformal field
$c_{p,q} (z, \zb)$.

Whereas $C_{p,q} \vert$ transforms homogenously under
conformal and $U(1)$-transformations,
the (bosonic) space-time field $[D, \dab ] C_{p,q}  \vert$
does not unless $p=0=q$.
In fact, $[D , \dab ] C_{p,q}$ is not a superconformal field.
The remedy to this problem consists of adding an appropriate term
to the projection operator $[D, \dab ]$ which eliminates
the unwanted contribution in the transformation law of
$[D, \dab ] C_{p,q}  \vert$
- see section 3.3.

\subsection{From $N=2$ to $N=1$ formalism}

Since $N=2$ theories
are often formulated in terms of $N=1$ superfields,
we will briefly outline how the latter are extracted from $N=2$ superfields.

A generic $N=2$ superfield (\ref{suf})
can be rewritten in terms of the odd coordinates
\begin{equation}
\label{nc}
\th_1  =  (\th + \tb )/2
\qquad , \qquad
\th_2  =   (\th - \tb )/2
\end{equation}
and the complex conjugate variables $\tb_1 = (\th_1 )^{\ast}, \,
\tb_2 = (\th_2 )^{\ast}$ which
give rise to the $N=1$ derivative
\begin{equation}
D_1  \equiv  {\pa \over \pa \th_1} + \th_1 \pa
=  D + \dab
\quad , \quad
(D_1)^2  =  \pa
\qquad (\, {\rm and} \ \, {\rm c.c.} \, )
\ \ .
\end{equation}
Indeed, one finds
\begin{equation}
\label{21}
C(z,\zb ,\th_1 , \tb_1  , \th_2 , \tb_2 )
= a + \th_2 \alpha + \tb_2 \beta + \th_2 \tb_2 b
\ \ ,
\end{equation}
where $a, \, \alpha , \, \beta , \, b$ represent 
$N=1$ superfields depending on the variables $z, \zb , \th_1 , \tb_1$. 
If $C$ is even, then $a, \,  b \ (\alpha , \, \beta )$ are even (odd).

\section{Derivatives, connections and covariant operators}

Along the lines of the $N=1$ theory \cite{cco},
we now introduce the necessary ingredients for the construction
and description of covariant operators.

\subsection{Schwarzian derivative}

The {\em super Schwarzian derivative} \cite{jc,cco} associated to
a superconformal change of coordinates $\uz \to \uz^{\pr} (\uz )$
is defined by
\begin{eqnarray}
-{\cal S} ({\uz} ^{\prime} , {\uz} )
& = & 2 \, {\rm e}^{-{1 \over 2} (w + \bar w)} \
[D,\dab]
\, {\rm e}^{{1 \over 2} (w + \bar w)}
\nonumber  \\
&=& [D,\dab] (w+ \bar w) +D(w+\bar w ) \dab (w+\bar w)
\nonumber  \\
&=& \pa (w -\bar w) +(D\bar w ) (\dab w)
\label{ssd}
\\
& = &
\frac{\pa \dab \tb^{\prime}}{\dab \tb^{\prime}}
\ - \
\frac{\pa D\th^{\prime}}{D \th^{\prime}}
\ + \
\frac{ \pa \tb^{\prime} }{ \dab \tb^{\prime} }
\; \frac{ \pa \th^{\prime} }{ D \th^{\prime} }
\ \ .
\nonumber
\end{eqnarray}
It satisfies the `chain rule'
\begin{equation}
\label{84}
{\cal S} ( {\uz} ^{\prime \prime} , {\uz} ) \ = \
{\rm e}^{ - ( w + \bar w)}\,
{\cal S} ( \uz ^{\prime \prime} , \uz ^{\prime} ) \ + \
{\cal S} ( \uz  ^{\prime} , \uz )
\ \ .
\end{equation}
By using equations (\ref{ssd}) and (\ref{w}), we can extract the bosonic
component fields of ${\cal S}$:
\begin{eqnarray}
{\cal S} ( \uz ^{\pr} , \uz ) \vert & = & \pa k
\\
{[D,\dab ]} {\cal S} (\uz ^{\pr} , \uz ) \vert & = &
S (z^{\prime} , z) \, + \, {1 \over 2} \, (\pa k)^2
\ \ .
\nonumber
\end{eqnarray}
Here,
$S ( z^{\prime} , z)$
denotes the ordinary
Schwarzian derivative (\ref{int}): it represents the `conformal
part' of ${\cal S}$ while $\pa k$ represents its `$U(1)$ part'.

\subsection{Projective coordinates}

A {\em superprojective (super M\"obius)
mapping} is a superconformal change
of coordinates
${\UZ} = (Z, \TH , \TB ) \rightarrow \UZ^{\pr} (\UZ) $ satisfying
${\cal S} (\UZ^{\pr} , \UZ )  = 0$.
By virtue of the definition (\ref{ssd}), this condition is equivalent to the relation
\begin{equation}
\label{S=0 bis}
(D_{\TH}
\TH^{\pr} )(D_{\TB} \TB^{\pr} ) \ =\ (cZ+d+\TH \gb +\TB \ga )^{-2}
\ \ ,
\end{equation}
where $c$ and $d$ ($\ga$ and $\gb$) are Grassmann even (odd) constants.
The latter relation can be integrated so as to obtain
the general expression of a superprojective mapping
$\UZ \to \UZ^{\pr}$:
by taking into account the conditions
$D_{\TH} \TB^{\prime} =0=
D_{\TB} \TH^{\prime}$ and
$
D_{\TH} Z^{\prime} = {1 \over 2} \, \TB^{\prime} \, D_{\TH} \TH^{\prime}
, \,
D_{\TB} Z^{\prime} = {1 \over 2} \, \TH^{\prime} \, D_{\TB} \TB^{\prime}
$,
one finds
\begin{eqnarray}
\TH^{\pr} (\UZ ) \! & \! = \! & \!
{\rm e}^{+h} \ \{ -2 \frac{\al Z+\be }{cZ+d}
 \, +\, {\TH \over
cZ+d}  \left[ 1+2  \, \frac{\ab Z+ \bb}{cZ+d} \, \ga \right] \, -\,
{\TH \TB \over (cZ+d)^2 } \, \ga \}
\\
\TB^{\pr} (\UZ ) \! & \! = \! & \!
{\rm e}^{-\bar h} \ \{ -2 \frac{\ab Z+\bb
 }{cZ+d} \, +\,
{\TB \over cZ+d} \left[ 1-2 \, \frac{\al Z+ \be}{cZ+d} \, \gb \right]
\, + \, {\TH \TB \over (cZ+d)^2 } \, \gb \}
\nonumber
\\
Z^{\prime} ( \UZ ) \! & \! = \! & \!
\frac{aZ+b}{cZ+d} \
\left[ 1+2(\al \bb +\ab \be ) + 8 \al \bb \ab \be \right]
\nonumber
\\
& &
- \frac{ \TH (\ab Z+\bb )
+ \TB ( \al Z+\be  )
 }{(cZ+d)^2}
\left[ 1
+2(\al \bb +\ab \be )  \right]
 +
\TH \TB \, \pa \left[ \frac{\al Z+\be }{cZ+d} \; \frac{\ab Z+\bb }{cZ+d}
\right]
.
\nonumber
\end{eqnarray}
Here $h,\bar{h},a,b$ ($\al, \be , \ab, \bb$) are Grassmann even (odd)
constants which satisfy the relations
\begin{eqnarray}
\al d-\be c  &= & \ga
\quad , \quad
\ \ \ \ \ \ ad - bc \; = \; 1
\\
\ab d-\bb c & = & \gb
\quad , \quad
2(\al \bb +\ab \be )  \; = \; h - \bar{h}
\ \ .
\nonumber
\end{eqnarray}
The lowest order component of $Z^{\prime} ( \UZ )$ represents an
ordinary projective transformation:
$Z^{\prime} \vert =
(a Z | + b)(cZ| +d)^{-1}$.

A {\em superprojective structure} on a SRS is an atlas of local coordinates for
which all changes of charts
$\UZ \rightarrow \UZ^{\prime}$ are superprojective mappings.

A {\em quasi-primary superfield}
${\cal C}_{p,q} (\UZ, \bar{\UZ})$
of superconformal weight $(p,q)$
transforms according to the rule
\begin{equation}
\label{rul}
{\cal C}^{\pr} _{p,q} ({\UZ} ^{\pr} , \bar{\UZ}
^{\pr} ) =
 (cZ+d+\TH \gb +\TB \ga )^{p+q}
\ {\rm e}^{-(p-q) H}
\ {\cal C}_{p,q} (\UZ , \bar{\UZ} )
\end{equation}
with
\[
H ( \UZ) = h + \ds{1 \over cZ+d}
\left[ 2 (\ab Z + \bb ) + \TH \gb - \TB \ga +{1\over 2}
\TH \TB c \right]
\]
under a superprojective change of coordinates $\UZ \to \UZ^{\pr}$.

\subsection{Projective connection}

A {\em superprojective (super Schwarzian) connection} 
on a SRS is a collection
${\cal R} \equiv {\cal R}_{\th \tb} (\uz)$ of superfields (one for each
coordinate
patch) which are locally superanalytic (i.e. $D_- {\cal R} =0=
\dab_- {\cal R}$)
and which transform
under a superconformal change of coordinates
according to
\begin{equation}
\label{rtr}
{\cal R}^{\prime}
({\uz} ^{\prime} )
= {\rm e}^{w+\bar w} \left[ {\cal R} ({\uz} )
- {\cal S} ({\uz} ^{\prime} , {\uz} )
\right]
\ \ .
\end{equation}
The existence of such connections on compact SRS's of arbitrary
genus can be proven along the lines of the $N=0$ and $N=1$
theories \cite{cco}.

If ${\bf Z}$ belongs to a projective atlas, then the
super Schwarzian derivative of the
superconformal mapping $\uz \rightarrow \UZ$ represents
a projective connection,
\begin{equation}
\label{ex}
{\cal R} ( \uz ) = {\cal S} ( \UZ  ,  \uz )
\ \ .
\end{equation}
In fact,
equation (\ref{84}) then implies the transformation law (\ref{rtr})
and it ensures that the expression (\ref{ex}) is inert under a
projective change of coordinates
$\UZ \rightarrow \UZ^{\prime}$.
The relation (\ref{ex}) establishes the equivalence between
projective structures
and projective connections.

The superfield $\rr$ admits the $\th$-expansion
\[
\rr_{\th \tb} ( \uz ) =  \rho_z
+\th \eta_{\th z}
+\tb \bar{\eta}_{\tb z}
+ \th \tb [ - {1 \over 2} r_{zz} ]
\ \  .
\]
The component fields $\rho = \rr  \vert, ...,
r = [D,\dab ] \rr  \vert$
are locally
holomorphic ($0 = \pab \rho = ... = \pab r$) and
the bosonic components have the
transformation laws
\begin{eqnarray}
\label{26}
\rho^{\pr} & = & (\pa z^{\pr} )^{-1} [ \, \rho - \pa k  \, ]
\\
r^{\pr} & = & (\pa z^{\pr} )^{-2} [ \, r - S -  (\rho -
{1 \over 2}  \pa k ) \pa k \, ]
\ \ .
\nonumber
\end{eqnarray}
From these equations,
we conclude that the quantity $R = r - {1 \over 2}
\rho^2 $ transforms like (\ref{int}), i.e. like an ordinary projective
connection while $\rho$ represents a $U(1)$-connection.

Using a superprojective connection,
we can define the second order operator
\begin{equation}
\label{qua}
{\cal D}_{p,q} =
p \dab D - q D \dab + pq \rr
\end{equation}
which projects out an
ordinary conformal field from the superconformal field $C_{p,q}$:
\[
\left( {\cal D}_{p,q} C_{p,q} \right)^{\pr}  \! \vert \; = \;
{\rm e}^{-{k \over 2} (p-q)} \
(\pa z^{\pr} )^{- {1 \over 2} (p+q+2)}
\left( {\cal D}_{p,q} C_{p,q} \right) \! \vert
\ \ .
\]

\subsection{Affine connection, covariant derivative and Miura
transformation}

A {\em superaffine connection} on a SRS is a collection
$B \equiv B_{\th} (\uz), \bar B \equiv \bar B _{\tb} (\uz)$
of superfields (one for each coordinate patch) which are
locally superanalytic, which satisfy the chirality conditions
$DB = 0 = \dab \bar B$ and which
transform
under a superconformal change of coordinates
according to
\begin{eqnarray}
B^{\prime}
({\uz} ^{\prime} )
& = & {\rm e}^{w} \left[ B ({\uz} )
+ D  \bar w \right]
\nonumber
\\
\bar B ^{\prime}
({\uz} ^{\prime} )
& = & {\rm e}^{\bar w} \left[ \bar B ({\uz} )
+ \dab    w \right]
\ \ .
\end{eqnarray}
The only compact SRS's which admit a globally defined affine
connection are those of genus one \cite{cco}.
Nevertheless, these quantities can always be introduced
locally and they represent extremely useful computational tools
for deriving glo\-bally well-defined results.

Using an affine connection, we can introduce
{\em supercovariant derivatives}
\begin{eqnarray}
\label{cd}
\nabla_{p,q}  =  D - q B & : & {\cal F}_{p,q} \ \longrightarrow
{\cal F}_{p+1,q}
\\
\bar{\nabla}_{p,q}  =  \dab  - p \bar B & : &
{\cal F}_{p,q} \ \longrightarrow
{\cal F}_{p,q+1}
\ \ .
\nonumber
\end{eqnarray}
Products of covariant derivatives are defined by taking
into account the weights, e.g.
\[
\nabla \bar{\nabla}  \nabla  C_{p,q}
= \nabla_{p+1,q+1} \bar{\nabla} _{p+1,q} \nabla_{p,q}C_{p,q}
\ \ .
\]
From the very definitions, it then follows that
\begin{equation}
\nabla^2 = 0 = \bar{\nabla}^2
\ \ .
\end{equation}

Locally, we can express $B$ and $\bar B$ as
\begin{equation}
\label{fac}
B   =  DQ
\quad , \quad
\bar B  =  \dab Q
\end{equation}
with
\begin{equation}
\label{qq}
Q \; = \;
{\rm ln} \, \left( D\TH \,
\dab \TB\right)
\; = \; \ln \, \left( \pa Z + {1 \over 2} \, \TH \pa \TB +
{1 \over 2} \, \TB \pa \TH\right)
\ \ ,
\end{equation}
where
$\TH$ and $\TB$ belong to an atlas of superprojective coordinates
$\UZ$.
Then, we have the operatorial relations
\begin{eqnarray}
\nabla_{p,q} & = & (\dab \TB )^q \cdot D \cdot (\dab \TB)^{-q}
\\
\bar{\nabla} _{p,q} & = &
(D\TH )^p \cdot \dab \cdot ( D\TH)^{-p}
\ \ .
\nonumber
\end{eqnarray}

Affine and projective connections are related by
the {\em super Miura
transformation}
\begin{eqnarray}
\label{miu}
{\cal R}
& = & D \bar{B} -  \dab B  - B \bar B
\\
& = & [D,\dab] Q -(DQ)( \dab Q)
\ \ .
\nonumber
\end{eqnarray}
This formula
follows from the expression of the projective connection
in terms of the Schwarzian derivative
(i.e equation (\ref{ex})) or, equivalently,
by comparing the expressions
(\ref{sb}) and (\ref{ff}) below for
the basic conformally covariant operator
${\cal L}_1^{sym}$.

From eq.(\ref{miu}) we can determine some explicit expressions for
the component fields
$[D, \dab ] \rr \vert = r = R + {1 \over 2} \rho^2$ and
$\rr \vert = \rho$. In fact, by taking into account eq.(\ref{qq}), one
recovers the ordinary Miura transformation for the projective
connection $R$ and an explicit expression for $\rho$:
\begin{eqnarray}
R & = & \pa ^2 \ln \, \pa Z\vert \, - \, {1 \over 2} \,
(\pa \ln \, \pa Z\vert )^2
\\
\rho & = & \left( [ D,\dab ]\ln \, \pa Z \right) \! \vert
\ \ .
\nonumber
\end{eqnarray}
Furthermore, from equations
(\ref{fac}) and (\ref{qq}),  we conclude that
\begin{eqnarray}
b & \equiv & \dab B \vert = {1 \over 2} \,
(\pa \ln \, \pa Z\vert  - \rho )
\\
\bar b & \equiv & D \bar B \vert ={1 \over 2} \,
(\pa \ln \, \pa Z \vert + \rho)
\ \ .
\nonumber
\end{eqnarray}
From these expressions, we obtain the {\em conformally
and} $U(1)$-{\em covariant space-time derivative}
\begin{equation}
\{ \nabla , \bar{\nabla} \} \, C_{p,q} \vert \, = \,
\left[  \, \tilde{\pa} _{p,q} -\ds{p+q \over 2}
\, (\pa \, {\rm ln} \, \pa Z \vert
) \, \right] \, c_{p,q}
\quad \quad {\rm with} \quad
\tilde{\pa}_{p,q} \equiv \pa - {p-q \over 2} \rho
\ \ .
\label{uco}
\end{equation}
Here,
$\tilde{\pa}_{p,q}$ denotes the
$U(1)$-covariant derivative associated to the
$U(1)$-connection $\rho$ (cf. eq.(\ref{26})) and
the local expression
$\pa \, {\rm ln} \, \pa Z \vert$
transforms like an affine connection with respect to a conformal
change of coordinates $z \to z^{\prime} (z)$ \cite{cco}.

\subsection{Determination of conformally covariant operators}

We are interested in superconformally covariant differential operators
which are globally defined on compact SRS's of any genus.
In order to construct these quantities, we start from an
operator ${\cal L}$ which is a polynomial in
the covariant derivatives $\nabla, \bar{\nabla}$ and require
that it only depends on the affine connections $B, \bar{B}$
through the particular combination $\rr = D\bar B - \dab B - B\bar B$
(which is globally defined):
using
a variational argument \cite{sch, dfiz},
one imposes $\delta {\cal L} =0$ while varying $B, \bar B$
subject to the condition that $\rr$ is fixed,
\[
0 = \delta \rr =
D\delta \bar B - \dab \delta B - \delta B \bar B - B \delta \bar B
\ \ .
\]
The latter relation can be rewritten as
\begin{equation}
\label{dr}
\nabla \delta \bar B  = \bar{\nabla} \delta B
\ \ ,
\end{equation}
and the nilpotency of $\nabla, \bar{\nabla}$ then yields
\begin{equation}
\bar{\nabla} \nabla \delta \bar B = 0 = \nabla \bar{\nabla} \delta B
\ \ ,
\end{equation}
while the chirality properties of $B, \bar B$ imply
\begin{equation}
\label{dc}
\nabla \delta B  = 0 =  \bar{\nabla} \delta \bar B
\ \ .
\end{equation}

There are two basic classes of covariant differential operators:
besides the
super Bol operators (which only depend
on a projective
structure or equivalently on a projective connection), one can
introduce operators which also depend linearly
on a conformal field.
These two classes will be determined in sections 4 and 5, respectively,
and at the end of section 5, we will show that the most
general covariant operator can be written as a sum of these.

\section{Super Bol operators}

We successively discuss the cases where the differential
operator ${\cal L}$ contains
an even and odd number of derivatives $\nabla, \bar{\nabla}$.
All operators discussed in this section
are supposed to be {\em normalized} in the sense
that the coefficients of the highest order derivatives are constant.

\subsection{Even
number of derivatives}

The general form of a normalized covariant operator
involving an even number of derivatives is
\begin{equation}
\label{evn}
{\cal L}_n^{even} =
\alpha (\nabla \bar{\nabla})^n + \beta (\bar{\nabla} \nabla)^n
\qquad ( \,
n \in {\bf N}^{\ast}  \ ,  \
(\alpha , \beta) \neq (0,0) \, )
\end{equation}
and such that
\begin{equation}
{\cal L}_n^{even} \ : \ {\cal F}_{p,q} \ \longrightarrow \
{\cal F}_{p+n,q+n}
\qquad {\rm for} \ \, {\rm some} \ \, (p,q)
\ \ .
\end{equation}
Thus, the operator does not modify the $U(1)$ charge of the fields.
In order to simplify the notation, we have not spelled out the dependance
of ${\cal L}$ on $p$ and $q$.

Imposing $\delta {\cal L}^{even} _n =0$
and using eqs.(\ref{dr})-(\ref{dc}),
one finds that $p,q$ and $n$ are related:
the general solution reads
\begin{eqnarray}
{\cal L}_n^{even}  & = & q (\nabla \bar{\nabla})^n
-p (\bar{\nabla} \nabla )^n
\qquad \qquad {\rm with} \quad p+q = -n
\nonumber
\\
& = & \{ \nabla , \bar{\nabla} \} ^{n-1}
(q  \nabla \bar{\nabla}
-p \bar{\nabla} \nabla )
\ \ .
\label{res}
\end{eqnarray}
Here and in the following, the constant
overall factor of ${\cal L}$ has been chosen in a convenient way.
For $n=1$ and $n=2$, we have the explicit expressions
\begin{eqnarray}
{\cal L}_1^{even}
\! & \! = \! &\!
q D\dab - p \dab D - pq \rr
\qquad \qquad \qquad\qquad
\quad \qquad \quad \ \ {\rm with} \quad p+q =-1
\nonumber
\\
\! & \! = \! & \!
{p+q \over 2} \, {\cal L}_1^{sym}
- {p-q \over 2} \left( \pa - {p-q  \over 2} \rr \right)
\label{gs}
\\
{\cal L}_2^{even}
\! & \! = \! & \!
{p+q \over 2}
\, {\cal L}_2^{sym} \; - \;
{p-q \over 2} \left(
\pa^2 - \rr [D , \dab ] - 2\,
{p-q \over 2} \rr \pa
+ (\dab \rr ) D
- (D \rr ) \dab
\right.
\nonumber
\\
& & \qquad \qquad \qquad \qquad  \qquad \quad
\left.
- {p-q \over 2} \, \pa \rr -pq \rr ^2 \right)
\quad {\rm with} \quad p+q =-2
\ ,
\nonumber
\end{eqnarray}
where ${\cal L}_1^{sym}$ and
${\cal L}_2^{sym}$ denote the symmetric solution ($p=q$),
see eq.(\ref{sb}) below.
Thus, the covariant operator ${\cal L}_n^{even}$ has a conformal part
(a multiple of ${\cal L}_n^{sym}$)
and a $U(1)$ part, each of which is proportional to the
corresponding weight.

Projection to components is achieved by virtue of the operator
${\cal D}_{p,q}$
introduced in eq.(\ref{qua}):
\begin{equation}
\label{pc}
\left( {\cal D}_{p+n,q+n}
\, {\cal L}_n^{even}
\, C_{ p,q } \right) \! \vert
\ = \ pq \, L_{n+1} ^{p,q} c_{p,q}
\ \ .
\end{equation}
Here,
$L_n^{p,q}$ denotes the {\em generalization of the
usual Bol operator} $L_n$ (depending on the
ordinary projective connection
$R = r - {1\over 2}
\rho^2$)
{\em to charged conformal fields}:
it amounts to replacing the ordinary partial derivative (acting on
$c_{p,q}$) by the
$U(1)$-covariant derivative
$\tilde{\pa}_{p,q}$ introduced in eq.(\ref{uco}), e.g.
\begin{equation}
\label{chb}
\begin{array}{llll}
L_2^{p,q} & = &
\tilde{\pa}_{p,q} ^2 + {1 \over 2} \, R
&\quad  {\rm with} \ \, p+q =-1
\\
L_3^{p,q} & = &
\tilde{\pa}_{p,q} ^3 + 2 R \tilde{\pa}_{p,q} + (\pa R )
&\quad {\rm with} \ \, p+q =-2
\ \ .
\end{array}
\end{equation}

Obviously, the general solution (\ref{res}) involves
two  different classes
of solutions: the
{\em symmetric solution} ($p =q$)
for which the operator acts on neutral fields
and the {\em asymmetric solutions} ($p\neq q$)
among which we find as particular cases the {\em chiral} ($p=0$) and
{\em anti-chiral} ($q=0$) {\em solutions}.

\subsubsection{Symmetric solution}

The symmetric solution
${\cal L}_n ^{sym} : {\cal F}_{-{n \over 2}, -{n \over 2}} \to
{\cal F}_{{n \over 2}, {n \over 2}}$
has the general form
\begin{eqnarray}
\label{ff}
{\cal L}_n ^{sym} & = &
( \nabla \bar{\nabla} )^n - (\bar{\nabla} \nabla )^n
\\
& = & \pa^{n-1} [D , \dab] + {n \over 2} \, \rr \pa^{n-1} + ...
\nonumber
\end{eqnarray}
and the simplest examples are
\begin{eqnarray}
{\cal L}_1^{sym} \! \! & \!  \! = \! \!  & \! 
[D,\dab ] + {1\over 2} {\cal R}
\nonumber
\\
\label{sb}
{\cal L}_2^{sym} \! \! & \! \! =  \! \!&  \! \pa [D,\dab ]
+ {\cal R} \pa
- (D{\cal R} ) \dab
- (\dab {\cal R} ) D
+ (\pa {\cal R})
\\
{\cal L}_3^{sym} \! \! & \! \! = \! \! &  \! \pa^2 [D,\dab ]
+ {3 \over 2} \rr \pa^2
- 3 (D{\cal R} ) \pa \dab
- 3 (\dab {\cal R} ) \pa D
+ 3 (\pa {\cal R})\pa
\nonumber
\\
& & +
\left( {1 \over 2} [ D, \dab ] \rr - {1 \over 4} \rr^2 \right) [D, \dab ]
- \left( 2 \pa D\rr + {1 \over 4} D \rr^2 \right) \dab
- \left( 2 \pa \dab \rr - {1 \over 4} \dab \rr^2 \right) D
\nonumber
\\
& &
+ {3\over 2} \left( \pa^2 \rr + {1 \over 2} \rr  [D, \dab ]\rr
- {1\over 4} \rr^3 -2 (D \rr)(\dab \rr) \right)
\nonumber
\\
{\cal L}_4^{sym}  \! \! &  \! \! =
\! \! &  \! \pa ^3 [ D,\dab  ] +
2 \rr \pa ^3 -6 ( \dab \rr ) \pa ^2 D
-6 (D\rr ) \pa ^2 \dab +6 \pa \rr \pa ^2 
  \nonumber \\
\nonumber &  & 
 +
\left( 2 ([ D,\dab ] \rr ) - \rr ^2 \right) \pa [ D,\dab ]
-
\left( 8 \pa \dab \rr - 2 \rr \dab \rr \right) \pa D -
\left( 8 \pa D \rr +2\rr D \rr \right) \pa \dab \\
\nonumber & & +
\left( 6 \pa ^2 \rr +4\rr [ D,\dab ]  \rr
-15 D\rr \dab \rr -2\rr ^3 \right) \pa +
\left( \pa [ D,\dab ] \rr -\rr \pa \rr \right) [ D,\dab ]  \\
\nonumber & & -
\left(3 \pa ^2 \dab \rr -\rr \pa \dab \rr
-\frac{3}{2} \dab \rr \pa \rr +
\frac{9}{2} \dab \rr [ D,\dab ] \rr -2 \rr ^2 \dab \rr \right) D\\
\nonumber & &  -
\left( 3 \pa ^2 D\rr +\rr \pa D\rr +\frac{3}{2} D \rr \pa \rr +
\frac{9}{2} D\rr [ D,\dab ] \rr -2 \rr ^2  D\rr \right) \dab \\
\nonumber & &  +
2\rr \pa [ D,\dab ]  \rr -10D\rr \pa \dab \rr +
10 \dab \rr \pa D \rr +4 \pa \rr [ D,\dab ]  \rr -
4 \rr ^2 \pa \rr +2 \pa ^3 \rr ,
\end{eqnarray}
where the derivatives only act on the field to their
immediate right.
As indicated above, the result for ${\cal L} _1 ^{sym}$ is nothing but the Miura
transformation (\ref{miu}). The operator ${\cal L}_n^{sym}$ has the hermiticity property
\[
\int dz d\th d\tb 
\, f \, {\cal L}_n ^{sym} g \, = \,  
(-1)^{n+1} 
\int dz d\th d\tb 
\, ({\cal L}_n ^{sym} f )\, g 
\qquad {\rm for} \ \; f , g \in 
{\cal F}_{-{n \over 2}, -{n \over 2}}
\ \ . 
\] 
The operators (\ref{sb}) play the role of Poisson (Hamiltonian)
operators in the description of superintegrable models -
see section 7.

\subsubsection{Chiral and anti-chiral solutions}

The anti-chiral ($q=0$) and
chiral ($p=0$) solutions read
\begin{eqnarray}
\label{chj}
{\cal L}_n^{anti} & = &  (\bar{\nabla} \nabla )^n
\ = \ \dab
(\nabla \bar{\nabla} )^{n-1}  D \ : \
{\cal F}_{-n,0} \to
{\cal F}_{0,n}^a
\\
{\cal L}_n^{chir} & = & (\nabla  \bar{\nabla} )^n
\ = \  D
(\bar{\nabla} \nabla )^{n-1} \dab \ : \
{\cal F}_{0,-n} \to
{\cal F}_{n,0}^c
\ \ .
\nonumber
\end{eqnarray}
For these operators, the target spaces are the anti-chiral
and chiral subspaces, respectively, since
$\dab {\cal L}_n^{anti} =0$ and $D{\cal L}_n^{chir} =0$.
Both classes of operators are related by hermitian conjugation:
\[
\int dz d\th d\tb 
\, f \, {\cal L}_n ^{chir} g \, = 
\, \int dz d\th d\tb 
\, ({\cal L}_n ^{anti} f ) \, g 
\qquad {\rm for} \ \; f \in 
{\cal F}_{-n,0} \ ,\  g \in 
{\cal F}_{0,-n} 
\ \ . 
\]
Differential operators of this type have been
introduced in reference \cite{pop}: they occur as Lax operators in the
$N=2$ supersymmetric KdV and KP hierarchies \cite{m, pop, dg}
 - see
section 5.

Since $q$ and $p$ vanish for
${\cal L}_n^{anti}$ and
${\cal L}_n^{chir}$, respectively, the
r.h.s. of eq.({\ref{pc}) vanishes for both of
these operators:
we will discuss their relation with 
ordinary Bol operators in the next section.

\subsection{Sandwich operators}

Before considering operators involving an odd number of derivatives,
it is worthwile to have another look at the chiral and anti-chiral
solutions.

The anti-chiral solution (\ref{chj}) has the form
\begin{equation}
{\cal L}_n^{anti}
C_{-n,0} =
\dab (\nabla \bar{\nabla} )^{n-1}  D
C_{-n,0}
\equiv
\dab {\cal K}_{n-1}  D
C_{-n,0}
\ \ ,
\end{equation}
where ${\cal K}_n$ will be referred to as `{\em sandwich operator}' for
obvious reasons.
Since $DC_{-n,0} \equiv
\Phi_{-n+1, 0}$ represents a chiral field,
we have obtained
a new conformally  covariant operator
\begin{equation}
\label{ncco}
\dab {\cal K}_{n-1} = \dab (\nabla \bar{\nabla})^{n-1}
\ : \ {\cal F}_{-n+1,0}^c \ \longrightarrow \
{\cal F}_{0,n}^a
\ \ .
\end{equation}
Although ${\cal K}_{n-1}$ formally coincides with
${\cal L}_{n-1}^{chir} = (\nabla \bar{\nabla})^{n-1}$, it does
not represent the same operator,
because it does not act on the same space.
We have the operatorial relations
\begin{eqnarray}
\label{sand}
\dab {\cal K}_1 & = & \dab [ D \dab + \rr ]
\ = \ \dab [ \pa + \rr ]
\\
\dab {\cal K}_2 & = & \dab [ \pa ^2
+3  \rr \pa + ( \dab D \rr ) + 2 ( D \dab \rr )  + 2 \rr^2 ]
\ \ .
\nonumber
\end{eqnarray}
Note that the operator ${\cal K}_n$ can be written in different
forms since it is only defined as `sandwiched' between $\dab$ and $D$.

The projection to component field expressions is simply done
by applying $D$ to $\dab {\cal K}_n$:
\begin{equation}
\label{met}
(D \dab {\cal K}_n C) \vert \ = \ L_{n+1} ^{-n,0} \, c
\qquad {\rm for} \quad C \in {\cal F}_{-n,0}^c
\ \ .
\end{equation}
Here, $C \, \vert = c$ and
$L_n^{p,q}$ denotes the generalization of the Bol operator $L_n$
to charged conformal fields -- see eq.(\ref{chb}).
This shows that ${\cal L}_n^{anti}$ also represents a supersymmetric
generalization of the Bol operators, though the latter
have to be projected out according to the procedure (\ref{met})
rather than (\ref{pc}).

Of course, one can apply the same line of reasoning to
the chiral solution for which
\begin{equation}
{\cal L}_n^{chir}
C_{0,-n} =
D  (\bar{\nabla} \nabla)^{n-1}  \dab
C_{0,-n}
\equiv
D  \bar{{\cal K}} _{n-1} \dab
C_{0,-n}
\end{equation}
and
\begin{equation}
D \bar{{\cal K}} _{n-1} \ : \ {\cal F}_{0,-n+1}^a \ \longrightarrow \
{\cal F}_{n,0}^c
\ \ .
\end{equation}

\subsection{Odd number of derivatives}

The general form of a normalized covariant operator
involving an odd number of derivatives is
\begin{eqnarray}
\label{od1}
{\cal L}_n^{odd} & = &
\bar{\nabla} (\nabla \bar{\nabla})^n
\\
{\rm or} \quad
\bar{{\cal L}} _n^{odd} &= &
\nabla (\bar{\nabla} \nabla)^n\qquad
\qquad {\rm with} \ \; n = 0,1,2,..
\nonumber
\end{eqnarray}
acting on ${\cal F}_{p,q}$.
These operators modify the $U(1)$ charge of the fields.

For $n=0$, the condition of covariance
$\delta {\cal L} =0$ leads to the explicit results
\begin{eqnarray}
{\cal L}_0^{odd} C_{0,q}
&=& \bar{\nabla} C_{0,q}
= \dab C_{0,q}
\\
\bar{{\cal L}} _0^{odd} C_{p,0}
&=& \nabla C_{p,0} =
D C_{p,0}
\ \ .
\nonumber
\end{eqnarray}
For $n\geq 1$, we find
\begin{equation}
\label{nncco}
{\cal L}_n^{odd} = \dab (\nabla \bar{\nabla} )^n
\ : \ \{ C \in {\cal F}_{-n,0} \ / \ {\cal L}_n^{anti} C =0 \}
\ \longrightarrow \ {\cal F}^a_{0, n+1}
\ \ ,
\end{equation}
the simplest examples being
\begin{eqnarray}
{\cal L}_1 ^{odd} C_{-1,0} & = &
\dab [ \pa  + \rr ]
C_{-1,0}
\\
{\cal L}_2 ^{odd} C_{-2,0} & = &
\dab [ \pa^2
+3  \rr \pa -(\dab \rr) D + ( \dab D \rr ) + 2 (D \dab \rr ) + 2 \rr^2 ]
C_{-2,0}
\nonumber
\end{eqnarray}
with
\begin{eqnarray}
0 & = & {\cal L}_1^{anti} C_{-1,0} = \dab D C_{-1,0}
\\
0&=& {\cal L}_2^{anti} C_{-2,0} = \dab [ \pa  + \rr  ] D C_{-2,0}
\ \ .
\nonumber
\end{eqnarray}

If $C_{-n,0}$ represents a chiral field, it automatically satisfies
${\cal L}_n^{anti} C_{-n,0} = 0$, henceforth
\[
{\cal F}_{-n,0}^c \subset \{ C\in {\cal F}_{-n,0} \ / \
{\cal L}_n^{anti} C = 0 \}
\ \ .
\]
By comparing equations (\ref{ncco}) and (\ref{nncco}), we conclude that
${\cal L}_n^{odd} = \dab (\nabla \bar{\nabla})^n$
is an extension (in the functional analytic sense 
\cite{rs}) of the
operator $\dab {\cal K}_{n}
= \dab (\nabla \bar{\nabla})^n$:
both operators differ by their
domain of definition and act in the same way on the smaller
domain, i.e. on ${\cal F}_{-n,0}^c$.
The Bol operator $L_{n+1} ^{-n,0}$
acting on $c = C \vert$
is extracted in the same manner from
${\cal L}_n^{odd} C$ and
$\dab {\cal K}_n C$, i.e. according to eq.(\ref{met}).

\subsection{Comments}

\subsubsection{Super Bol operators in projective coordinates}

The Bol operator $L_n$
is most directly obtained
from the M\"obius covariant operator
$\pa_Z^n$ (acting on quasi-primary fields of appropriate weight)
by going over from the projective coordinates $Z$ to a generic
system of conformal coordinates $z$ by a conformal
transformation (see \cite{cco} and references therein).
Proceeding in the same way in the $N=2$ supersymmetric case,
we start from the differential operator
$\pa_Z^{n-1} [ D_{\TH} , D_{\TB} ]
= (D_{\TH} D_{\TB} )^n -
(D_{\TB} D_{\TH} )^n$
which transforms homogenously
under a superprojective change of coordinates $\UZ \to \UZ^{\pr}$
when acting
on quasi-primary fields of appropriate weight (see eq.(\ref{rul})):
\begin{equation}
\left(
\pa_Z^{n-1} [ D_{\TH} , D_{\TB} ] \,
{\cal C}_{ -{n \over 2} , -{n \over 2} }  \right)^{\pr} =
(cZ+d+\TH \gb +\TB \ga )^n
\ \pa_Z^{n-1} [ D_{\TH} , D_{\TB} ] \,
{\cal C}_{ -{n \over 2} , -{n \over 2} }
 \ \ .
\end{equation}
By passing over from the projective coordinates $\UZ$
to generic conformal coordinates $\uz$ by a
superconformal transformation, the operator
$\pa_Z^{n-1} [ D_{\TH} , D_{\TB} ]$ becomes the
super Bol operator ${\cal L}_n^{sym}$:  we have
\begin{equation}
\label{62}
\pa_Z^{n-1} [ D_{\TH} , D_{\TB} ] \,
{\cal C}_{ -{n \over 2} , -{n \over 2} }  \equiv
(D\TH \, \dab \TB )^{- {n \over 2} }
\ {\cal L}_n^{sym}
C_{ -{n \over 2} , -{n \over 2} }
\ \ ,
\end{equation}
where ${\cal C}$ and $C$ are related by
\begin{equation}
{\cal C}_{ -{n \over 2} , -{n \over 2} } ({\UZ} , \bar{\UZ})
\equiv (D\TH \, \dab \TB )^{ {n \over 2} }
\ C_{ -{n \over 2} , -{n \over 2} } (\uz , \bar{\uz} )
\ \ .
\end{equation}
In operatorial form, equation (\ref{62}) reads
\begin{equation}
\label{64}
{\cal L}_n ^{sym}  =
(D\TH \, \dab \TB )^{ {n \over 2} }
\cdot  \pa_Z^{n-1} [ D_{\TH} , D_{\TB} ]
\cdot  (D\TH \, \dab \TB )^{ {n \over 2} }
\ \ ,
\end{equation}
which means that
${\cal L}_n ^{sym}$ represents
the superconformally covariant
version of the differential operator $\pa^{n-1} [D , \dab]$
acting on $U(1)$-neutral fields.

The same line of arguments applies to the sandwich operator ${\cal K}_n$
acting on chiral fields. In fact,
$D C_{-n,0} =0$ is equivalent to
$D_{\TH} \, {\cal C}_{-n,0} =0$
where ${\cal C}_{-n,0} \equiv (D\TH )^n  \, C_{-n,0}$
and for such fields one has
\begin{equation}
\label{bolproj}
D_{\TB} \pa_Z ^n \, {\cal C}_{-n,0} =
( \dab \TB )^{-(n+1)} \  \dab {\cal K}_n C_{-n,0}
\ \ .
\end{equation}

\subsubsection{Recursion and factorization relations}

The super Bol operators can be determined by using
a {\em recursion relation} \cite{cco}, e.g. if
$\psi \in {\cal F}_{-{1\over 2}, -{1 \over 2}}$
is assumed to be Grassmann even, then
\begin{equation}
\label{rec}
{\cal L}^{sym}_1 \psi =0  \ \ \Longrightarrow \ \ {\cal L}^{sym}_n
\psi^n =0
\ \ .
\end{equation}
Thus,
one can determine the explicit form of ${\cal L}_2^{sym}$
by applying $\pa [D, \dab ]$ to $\psi^2$ while using
${\cal L}_1^{sym} \psi =0$
(i.e. $[D, \dab ] \psi = -{1\over 2} {\cal R} \psi $) and so on for
${\cal L}_3^{sym},..$.
The proof of (\ref{rec}) consists of going over
to superprojective coordinates.

The operators $\dab {\cal K}_n$ can be determined in the same way:
if $\psi \in {\cal F}_{-1,0}^c$
is Grassmann even, then
\begin{equation}
\dab {\cal K}_1 \psi =0  \ \ \Longrightarrow \ \ \dab {\cal K}_n
\psi^n =0
\ \ .
\end{equation}

If the super Bol operators are defined by virtue of M\"obius
covariant operators as in equation (\ref{64}),
their expressions
in terms of covariant derivatives (eq.(\ref{ff}))
represent {\em factorization formulae}. The latter may then be proven
by going over to projective coordinates for which $\nabla$ and
$\bar{\nabla}$ reduce to $D_{\TH}$ and $D_{\TB}$, respectively
\cite{cco}.

\subsubsection{From $N=2$ to $N=1$ operators}

An $N=2$ superfield $F$ can be projected to an $N=1$ superfield
$\left. F  \right \vert _{N=1}$ by evaluating it at
$\th = \tb, \, \th^- = \tb^-$.
By virtue of this projection, one finds that the quantity 
$\rr_1 \equiv -{1 \over 2} \left. (D - \dab)
 \rr \right \vert _{N=1}$
transforms like an $N=1$ superprojective connection, i.e. 
according to 
$\rr_1 ^{\prime} = {\rm e}^{3W} ( \rr_1 - {\cal S}_1 )$
where $W \equiv \left. w \right \vert _{N=1} =
 \left. \bar{w} \right \vert _{N=1}$ and where ${\cal S}_1$ 
denotes the 
$N=1$ super Schwarzian derivative. 

Since sandwich operators act on chiral fields $C$, we have 
$\dab C = (D+\dab)C \equiv D_1 C$
where $D_1$ is the basic $N=1$ derivative 
introduced in section 2.6. 
For the simplest sandwich operator, we thus obtain 
the expression 
\begin{eqnarray}
\nonumber
\left. ( \dab {\cal K}_1 C ) \right \vert _{N=1} &=&
 \left. \left( [\pa \dab +(\dab \rr )+\rr \dab ] C \right)  \right \vert _{N=1} \\
&=&
\left. \left( [ D_1 ^3 + \rr _1 ] C +[\frac{1}{2} (D_1 \rr ) + \rr D_1 ] C \right)
 \right \vert _{N=1}
\ \ , 
\nonumber
\end{eqnarray}
which represents a $U(1)$-covariant version of the
 $N=1$ super Bol operator $D_1^3 + \rr_1$.

\section{Covariant operators involving conformal fields}

In this section,
we introduce covariant
operators
${\cal M}^{(n)}_{{\cal W}_{p,q}}$ which
depend on a projective connection ${\cal R}$ and, in a linear way,
on a superconformal field
${\cal W}_{p,q}$.
We restrict our attention to the case of sandwich operators
since these are the only ones for which we will be able to
give a matrix representation.

Thus, our goal is to
construct a covariant
operator
which acts between the same spaces as
$\dab {\cal K}_n$ (cf.eq.(\ref{ncco})):
\[
\bar{D}{\cal M}^{(n)}_{{\cal W}_{p,q}}\ :\  {\cal F}^c_{-n,0} \
\longrightarrow \ {\cal F}^a_{0,n+1}
\ \ .
\]
To do so, we proceed as before and first write
${\cal M}^{(n)}_{{\cal W}_{p,q}}$ as a polynomial
in the nilpotent operators
$\nabla$ and $\bar{\nabla}$.
It turns out that a covariant operator can only be
obtained in the case of
neutral superfields
${\cal W}_{k,k}$ of conformal
weight $k \in {\bf N}$ with $1\leq k\leq n$:
the corresponding operator reads
\begin{equation}
{\cal M}^{(n)}_{{\cal W}_{k,k}} \ = \ \sum_{l=0}^{n-k} \left\{\
a^{(n)}_{kl}\ [(\nab \nabar)^l {\cal W}_{k,k}] \; +
\; b^{(n)}_{kl}\ [(\nabar \nab)^l {{\cal W}_{k,k}}]\ \right\}
(\nab \nabar)^{n-k-l}
\end{equation}
with constant
coefficients
$a^{(n)}_{kl}$ and $b^{(n)}_{kl}$. 
The latter are determined by imposing the condition
$\delta \bar{D} {\cal M}^{(n)}_{{\cal W}_{k,k}}=0$ and by using
eqs.(\ref{dr})-(\ref{dc}).
This leads to the following result (which is
unique up to a global factor):
\begin{equation}
\begin{array}{ccc}
a^{(n)}_{kl}\ =\ \frac
{\left( \begin{array}{c} n-k \\ l \end{array} \right)
\left( \begin{array}{c} k+l \\ l \end{array} \right) }
{\left( \begin{array}{c} 2k+l \\ l \end{array} \right)}
&,&
b^{(n)}_{kl}\ =\ \frac
{\left( \begin{array}{c} n-k \\ l \end{array} \right)
\left( \begin{array}{c} k+l-1 \\ l \end{array} \right) }
{\left( \begin{array}{c} 2k+l \\ l \end{array} \right)}
\end{array}
\end{equation}
for $l=1,..,n-k$ and $a^{(n)}_{k0}\ +\ b^{(n)}_{k0}\ =\ 1$ .
For instance,
\begin{eqnarray}
{\cal M}^{(n)}_{{\cal W}_{n,n}}&=& {\cal W}_{n,n}
\nonumber \\
{\cal M}^{(n)}_{{\cal W}_{n-1,n-1}}& =& {\cal W}_{n-1,n-1} \pa +
\frac{n-1}{2n-1}(D{\cal W}_{n-1,n-1})\bar{D}+
\frac{n}{2n-1}(D\bar{D}{\cal W}_{n-1,n-1}) \nonumber\\
& &
+\frac{n^2}{2n-1}{\cal R}{\cal W}_{n-1,n-1}
\ \ ,
\end{eqnarray}
where we recall that sandwich operators can be cast into different
forms since they are only defined as sandwiched between $\dab$ and $D$.
Component field
results are recovered by
applying the derivative $D\ $ and subsequently
projecting to the lowest component:
\begin{equation}
(D\bar{D}{\cal M}^{(n)}_{{\cal W}_{k,k}}  C_{-n,0} )\vert \ =\
\left[ \tilde{M}^{(n+1)}_{w_k}  -
\frac{n+k+1}{2k(2k+1)}
\, \tilde{M}^{(n+1)}_{v_{k+1}} \right]  c_{-n,0}
\end{equation}
with $w_k  = {\cal W}_{k,k}\vert$ and $v_{k+1}  = ({\cal D}_{k,k} {\cal
 W}_{k,k}) \vert$ where
${\cal D}_{k,k}$ denotes the second order operator (\ref{qua}).
Here, $\tilde{M}^{(n)}_{w_k}$ is the $U(1)$-covariant generalization of the
operator
$M^{(n)}_{w_k}$ mentionned in the introduction - see eq.(\ref{mwb}).

The {\em most general} normalized
superconformally covariant sandwich operator of order $n$
has the form
\begin{eqnarray}
{\cal L}^{(n)} & :  &
{\cal F} ^c _{-n,0} \ \longrightarrow \ {\cal F} ^a _{0,n+1}
\label{mgso}
\\
& &
C_{-n,0} \longmapsto \
{\cal L}^{(n)}
C_{-n,0} =
\dab \left[ \pa^n + a_1^{(n)} \pa^{n-1}
+ a_2^{(n)} \pa^{n-2} +  ...
+ a_n^{(n)} \right]
C_{-n,0}
\nonumber
\end{eqnarray}
with even coefficient functions
$a^{(n)}_k ({\bf z})$
transforming in an appropriate way under superconformal
changes of coordinates - see references \cite{dfiz,cco} and
\cite{gt1} for the $N=0$ and $N=1$ supersymmetric theories,
respectively.
It can be expressed as a sum of the previously introduced operators,
\begin{equation}
\label{san}
{\cal L}^{(n)} \ =\ \bar{D} \left[
{\cal K}_n + \sum^n_{k=2}
{\cal M}^{(n)}_{{\cal W}_{k,k}} \right] \ = \
\dab \left[ \pa^n + {n(n+1) \over 2} \rr \pa^{n-1}
+  ... \right]
\ \ ,
\end{equation}
which means that it can be parametrized in terms of a
projective connection ${\cal R}$  and
$n-1$ neutral superconformal fields ${\cal W}_{2,2}, ...,
{\cal W}_{n,n}$. The relation between these fields and the
coefficients $a^{(n)}_k$ is invertible and given by
differential polynomials.

Operators of the form (\ref{mgso}) with
$C_{-n,0} = D \Phi$ occur as Lax operators, e.g.
${\cal L}^{(1)}$ and
${\cal L}^{(2)}$ give rise to the $N=2$ super KdV and 
Boussinesq equations, respectively \cite{pop, dg}.

\section{Matrix representation of covariant operators}

In this section, we discuss the matrix representation
of the sandwich operators (\ref{san}) while starting with the
particular case of the super Bol operator
\begin{equation}
\label{fso}
\dab {\cal K}_n = \dab (\nab \nabar )^n= \dab [D-nB] [\bar{D}+\bar{B}]
[D-(n-1)B] [\bar{D}+2\bar{B}]
\cdots [D-B] [\bar{D}+n\bar{B}]
.
\end{equation}
The two conformally covariant differential equations
\begin{equation}
\label{sca}
D f\ =\ 0 \ \ \ \ \ \ ,\ \ \ \ \ \dab {\cal K}_n \, f \  =\ 0
\qquad
(\, f\in {\cal F}_{-n,0} \, )
\end{equation}
are equivalent to two systems of first-order differential equations
which can be cast into matrix form:
\begin{equation}
\label{mf}
Q_n^{fac} \, \vec{f}\ =\ \vec{0} \qquad ,\qquad \bar{Q}_n^{fac} \, \vec f
\ =\ \vec{0}
\ \ .
\end{equation}
Here,
$\vec f  = (f_1 , f_2 , ..., f_{2n},f)^t$ and
$Q_n^{fac} , \, \bar{Q}_n^{fac}$ denote the $(2n+1)\times (2n+1)$
matrix operators
\begin{equation}
\label{matricefac}
Q_n^{fac}\ =\  -\mu_- \, +\, D {\bf{1}} \, -\, BH \ \ ,\ \
\bar Q_n^{fac}\ =\ -\bar{\mu}_- \, +\, \bar D {\bf{1}} \, - \bar B \bar H
\ \ ,
\end{equation}
where
$\mu_-, \bar{\mu}_-, H, \bar H$ belong to the superprincipal embedding
$sl(2 |1)_{{\rm pal}} \subset sl(n+1|n)$ of
the Lie superalgebra
$sl(2 | 1)$ into the Lie superalgebra $sl(n+1
|n)$ - see eqs.(\ref{ppal}) of appendix A.
We note that
the operators (\ref{matricefac}) can be rewritten in the form
\begin{equation}
\label{gaugeconnection}
 Q_n^{fac} \ =\ D {\bf{1}} \, -\, A^{fac}_{\th} \qquad , \qquad
\bar{Q}_n^{fac}\ =\ \bar{D} {\bf{1}} \, -\, A^{fac}_{\tb}
\ \ ,
\end{equation}
where
$A^{fac}_{\th}$ and  $A^{fac}_{\tb}$ are to be interpreted as components
of a
$sl(2 |1)_{{\rm pal}}$-valued connection
- a view-point which will be further developped in the next section.

$Q_n^{fac}$ and $\bar{Q}_n^{fac}$ describe the {\em factorized form}
(\ref{fso}) {\em of the super Bol operator}. One obtains
an equivalent matrix representation for this operator by conjugation
with a group element $N\in SL(2 | 1)_{\pal} \subset SL(n+1|n)$,
\begin{equation}
\label{gfr}
Q_n\ =\ {\hat N}^{-1} \cdot Q_n^{fac} \cdot N\ \ ,\ \
\bar{Q}_n\ =\ {\hat N}^{-1} \cdot\bar{Q}_n^{fac}\cdot N
\ \ ,
\end{equation}
where $N$ is an
upper triangular matrix
with entries $1$ on the diagonal
\cite{ds,gt1}.
(The systems of equations (\ref{mf})
then become
$Q_n  \vec{f}^{\prime} = \vec{0} , \,  \bar{Q}_n  \vec f^{\prime}
 = \vec{0}$ with
$\vec f ^{\prime} \equiv N^{-1} \vec f
= ( f_1^{\prime} , ..., f_{2n}^{\prime},f)$, i.e. the component $f$
of $\vec f$ is not modified, which explains why
$Q_n$ and $\bar{Q} _n$ still represent the scalar equations (\ref{sca}).)
The hatted group element $\hat N = {\rm exp} \, \hat M$ follows from
$N= {\rm exp} \, M$ by changing the sign of the
anticommuting part of the
Lie superalgebra element $M$ \cite{drs, gt1}.

By using the gauge freedom (\ref{gfr}),
we can find a form of $Q_n$ and $\bar{Q}_n$
which makes the dependence on the superprojective structure manifest:
\begin{eqnarray}
\label{matsv}
Q_n & = & -\mu_- \, +\, D {\bf{1}} \, -\,
\frac{1}{2} \, [ {\cal R}\bar{\mu}_+ \, + \, (D{\cal R}) E_{++} ]
\\
\bar{Q}_n & = & -\bar{\mu}_- \, +\, \bar{D} {\bf{1}} \, +\,
\frac{1}{2} \, [ {\cal R}\mu_+ \, + \, (\bar{D}{\cal R}) E_{++} ]
\ \ .
\nonumber
\end{eqnarray}
This matrix representation of $\dab {\cal K}_n$ in which $\mu _{\pm}, \,
\bar{\mu}_{\pm}$ and $E_{++} \equiv - \{ \mu_+ , \bar{\mu}_+ \}$
belong to the Lie superalgebra $sl(2|1)_{{\rm pal}}$
exhibits most clearly the underlying algebraic structure
which is due to the covariance with respect to super M\"obius
transformations.
By unravelling the matrix equations
$Q_n  \vec{f}^{\prime} = \vec{0} , \,  \bar{Q}_n  \vec f^{\prime}
 = \vec{0}$,
one recovers our results (\ref{sand}) for the operator
$\dab {\cal K}_n$.

The matrix representation (\ref{matsv})
can be extended to the sandwich operators
involving superconformal fields that we discussed in section 5:
\begin{eqnarray}
\label{matv}
Q_n &=& -\mu_- \, + \, D {\bf{1}} \, - \,
\frac{1}{2}
\, \sum^n_{i=1}
\, \left[ i{\cal V}_{i,i}\bar{\mu}_+ + (D{\cal V}_{i,i})E_{++} \right]
 E_{++}^{i-1} \\
\nonumber
\bar{Q}_n &=& -\bar{\mu}_- \, +\, \bar{D} {\bf{1}} \, +\,
\frac{1}{2}
\, \sum^n_{i=1}
\, \left[ i{\cal V}_{i,i} \mu_+ + (\bar{D}{\cal V}_{i,i})E_{++} \right]
 E_{++}^{i-1}
\ \ .
\end{eqnarray}
Here,  ${\cal V}_{1,1} \equiv {\cal R}$
and ${\cal V}_{i,i}$ is a superconformal field for $i \geq 2$.
The  ${\cal V}_{i,i}$ $(i=1,..,n)$ can be expressed
in terms of the projective
connection ${\cal R}$ and the conformal
fields ${\cal W}_{k,k}$ $(k=2,..,n)$ appearing in the expression
(\ref{san}) by means of
invertible relations involving differential polynomials.
Henceforth,
the expressions (\ref{matv}) provide a matrix representation
for the most general sandwich operator.
Thanks to this result, explicit expressions for covariant
operators can be derived in a straightforward way.

If the sandwich operators are parametrized in the form
(\ref{mgso}), their matrix realization reads
\begin{eqnarray}
\label{hor}
Q^{hor} _n &=& -\mu_- \, + \, D {\bf{1}}
\\
\bar{Q} ^{hor} _n &=& -\bar{\mu}_- \, +\, \bar{D} {\bf{1}} \, +\,
\sum^n_{i=1} \, \left[ \lambda_i a^{(n)}_i E_{1, 2i} \, + \,
\mu_i  (\dab a^{(n)}_i) E_{1, 2i+1} \right]
\ \ ,
\nonumber
\end{eqnarray}
where $E_{ij}$ are the matrices with entries
$(E_{ij})_{kl} = \delta_{ik} \delta_{jl}$ and
\[
\lambda_i = { (n-i)! \over n! \; (i-1)!} \qquad , \qquad
\mu_i = { (n-i)! \over n! \  i!}
\ \ .
\]
This gauge choice is referred to as the {\em horizontal gauge}
\cite{dm} while
(\ref{matsv},\ref{matv})
is known as the
$osp(1|2)$
{\em highest weight gauge}. In fact, as we will now verify, the
combination
$Q_n + \bar{Q} _n$ leads to the matrix representation of $N=1$
superconformally covariant differential operators in terms of
highest weight generators of  $osp(1|2)_{{\rm pal}} \subset sl(2|1)
_{\pal}$
(see appendix A for the algebraic details).

By adding the expressions
(\ref{matv}), we find
\[
Q_n + \bar{Q}_n \, = \,
- \kappa_- + (D + \dab ) {\bf 1} +
\frac{1}{2}
\, \sum^n_{i=1}
\, \left[
i{\cal V}_{i,i} M_{2i-1} - (D -\dab ) {\cal V}_{i,i}  M_{2i} \right]
\ \ .
\]
Here,
$\kappa_- \equiv \mu_- +\bar{\mu} _- \in
osp(1|2)_{{\rm pal}}$ and $M_k \equiv
(\mu_+ - \bar{\mu} _+ )^k $ are highest weight generators of
$osp(1|2)_{{\rm pal}}$ while $D + \dab = D_1$ denotes the basic
$N=1$ derivative. 
By projecting from $N=2$ to $N=1$ superfields
 along the lines of section 4.4.3, we  
get
\begin{equation}
\label{N=1}
\left. Q^{(N=1)}_n \equiv (Q_n + \bar Q _n ) \right \vert _{N=1} \, = \,
- \kappa_- + D_1  {\bf 1} +
\, \sum^{2n}_{k=1}
V_{k+1} M_k
\end{equation}
with
\[
\left. V_{2i} = {i \over 2} {\cal V}_{i,i} \right \vert _{N=1}
\qquad , \qquad
V_{2i+1} = \left. -{1 \over 2} (D- \dab){\cal V}_{i,i}\right \vert _{N=1}
\qquad (\, i = 1, .., n \, )
\ \ .
\]
The expression (\ref{N=1})
coincides with the
$N=1$ matrix
representation found in references
\cite{gt1, gt2} up to the
fact that it involves a different (though equivalent) realization
of the superprincipal embedding of
$osp(1|2)_{{\rm pal}}$ in $sl(n+1|n)$.
The same realization can be obtained by choosing $Q_n$ and $\bar Q _n$
along the lines of reference \cite{dm}:
\begin{eqnarray}
\label{matp}
 Q^{\prime} _n &=& - \tilde{\mu} _+ \, + \, D {\bf{1}} \, + \,
\frac{1}{2}
\, \sum^n_{i=1}
\, \left[ (D{\cal V}_{i,i}) \tilde{E} ^i _{--} -
{\cal V}_{i,i} [ \tilde{\mu} _+ , \tilde{E} ^i _{--}]  \right]
\\
\qb ^{\prime} _n &=& + \tilde{\mb} _+ \, + \, \bar{D} {\bf{1}} \, - \,
\frac{1}{2}
\, \sum^n_{i=1}
\, \left[ (\dab {\cal V}_{i,i}) \tilde{E} ^i _{--} +
{\cal V}_{i,i} [ \tilde{\mb} _+ , \tilde{E} ^i _{--} ] \right]
\nonumber
\ \ .
\end{eqnarray}
Here, the tilde denotes the transposed matrix and the common structure
of the results (\ref{matv}) and (\ref{matp}) becomes apparent
if one takes into account the identities
\begin{equation}
[ \mu_- , E^i_{++} ] = -i \mb_+ E^{i-1} _{++}
\qquad , \qquad
[ \mb_- , E^i_{++} ] = -i \mu_+ E^{i-1} _{++}
\ \ .
\end{equation}

\section{Ward identities and $W$-algebras from a
zero curvature condition}

The highest weight
matrix representation  (\ref{matv}) of covariant sandwich operators
has the form
\begin{equation}
 Q_n \ =\ D {\bf{1}} \, -\, A_{\th} \qquad , \qquad
\bar{Q} _n \ =\ \bar{D} {\bf{1}} \, -\, A_{\tb} \ ,
\end{equation}
where
$A_{\th}$ and $A_{\tb}$ can
be interpreted as the spinorial
components of a connection with values in the Lie superalgebra
$sl(2 | 1)_{{\rm pal}} \subset sl(n+1 | n)$.
The spinorial components can be supplemented with
spatial components $A_z$ and $A_{\zb}$ represented by generic
elements of $sl(n+1 | n)$.
Altogether these fields represent the fundamental variables for the 
superspace formulation of a $(2,0)$-supersymmetric 
gauge theory based on graded Lie algebras. (For the 
general framework of such theories, 
we refer to the appendix of \cite{gt2}.)

Within the $N=1$ superfield formalism, 
it has been shown \cite{gt2} that
classical super $W$-algebras can be constructed 
in the following way: in the $z$-sector, one considers 
the highest
weight gauge for the spinorial components of the connection
(i.e. eqs.(\ref{matv}) in the present case),
all entries of the
connection components $A_z, A_{\zb} ,A_{\th},...$
 are supposed to be smooth
superfields and zero curvature
conditions are imposed on the
connection.
We will now apply this procedure to the $N=2$ theory. 

The integrability conditions for the system of
differential equations
$\vec{0} = Q_n \vec f = \bar Q _n \vec f = (\pab - A_{\zb} )\vec f$
are
the {\em zero curvature
conditions}
\begin{eqnarray}
\nonumber
0&=& DA_{\th} -\hat{A} _{\th} A_{\th}  \qquad , \qquad
0\ =\ \bar{D}A_{\tb} -\hat{A} _{\tb} A_{\tb} \\
\nonumber
A_z &=& DA_{\tb} +\bar{D} A_{\th} -
\hat{A} _{\th} A_{\tb} -\hat{A}_{\tb} A_{\th}  \\
\label{zc3}
0&=& DA_{\zb} - \pab A_{\th} -A_{\th} A_{\zb} + \hat{A}_{\zb}  A_{\th}
\\
0&=& \bar{D} A_{\zb}
- \pab A_{\tb} -A_{\tb} A_{\zb} + \hat{A}_{\zb}  A_{\tb}
\ \ ,
\nonumber
\end{eqnarray}
where the hat again denotes the automorphism of the Lie superalgebra
which reverses the signs of all odd elements.
As may be verified explicitly, the expressions (\ref{matv}) (involving
smooth superfields ${\cal V}_{i,i}$) verify
the constraints of the first line.
The second line is simply a redefinition constraint which
determines the component $A_z$ in terms of the variables
$A_{\th}$ and $A_{\tb}$.
Finally, $A_{\th}$ and $A_{\tb}$ having the form (\ref{matv}),
the last two conditions represent constraints for the entries
of the matrix
$A_{\zb}$. Actually, most of these constraints are algebraic equations
which allow to express the entries of $A_{\zb}$ in terms
of other fields while the other ones represent partial differential
equations for the independent fields.
These differential equations are the Ward identities for the
super $W_{n+1}$-algebra associated to $sl(n+1 \mid n)$.

In the simplest case ($n=1$),
the matrix $A_{\zb}$ only contains
one independent superfield $H$ satisfying the differential equation
\begin{equation}
\label{ward1}
{\cal L}_2^{sym} H \ =\ \pab {\cal R} \ \ ,
\end{equation}
where
${\cal L}_2^{sym}$ denotes the symmetric super Bol operator (\ref{sb}).
This relation is nothing but
the $(2,0)$ {\em superconformal Ward identity} \cite{dgg},
the superfield $H \equiv H_{\zb}^{\ z}$ being interpreted as the Beltrami
 superfield (in the so-called restricted geometry)
and the superprojective connection ${\cal R}$
as the stress-energy tensor.
It expresses the superdiffeomorphism invariance of the generating
functional $Z_c [H]$ in superconformal field theory.

For $n=2$, $A_{\zb}$ contains
two independent superfields
$H \equiv H_{\zb}^{\ z}$ and $G \equiv G_{\zb}^{\ zz}$ which are
associated to ${\cal R}$ and ${\cal V} \equiv {\cal V}_{2,2}$,
respectively. These variables satisfy the
system of differential
equations
\begin{eqnarray}
\label{ward2}
\pab {\cal R}\! & \! = \! & \!
{\cal L}_2^{sym} H+
\left[ 2 {\cal V} \pa - (\bar{D} {\cal V} )D
- (D {\cal V} )\bar{D} +2(\pa {\cal V} ) \right] G \\
\nonumber
\pab {\cal V} \! & \! =\! & \!
{-1\over 16} \left[{\cal L}_4^{sym} + 4{\cal M}^{(4)sym}_{{\cal V}}
\right] G +\left[
2 {\cal V} \pa - (\bar{D} {\cal V}) D
- (D {\cal V} )\bar{D} +(\pa {\cal V} ) \right] H \ ,
\end{eqnarray}
where
${\cal M}^{(4)sym}_{{\cal V} }$
is the  superconformally covariant operator\footnote{This
operator is an example of a
covariant operator involving a
conformal superfield and  generalizing
the symmetric Bol operators of section 4.1.1. These operators
can be determined in a systematic way
along the lines of section 5.}
\begin{eqnarray}
\nonumber
{\cal M}^{(4)sym}_{\cal V} \! & \! = \! & \!
5 \vv \pa [ D,\dab ] +5 D\vv\pa\dab -5 \dab \vv \pa D +
\frac{5}{2} \pa \vv [ D,\dab ]   +
(\frac{3}{2} [ D,\dab ]   \vv +7 \rr \vv )\pa \\
\nonumber & & +
(3 \pa D \vv +\rr D\vv -9D\rr \vv )\dab -
(3 \pa \dab \vv -\rr \dab \vv +9\dab \rr \vv )D \\
\nonumber & & +
\pa [ D,\dab ]   \vv +3\rr \pa \vv -
5\dab \rr D\vv -5D\rr \dab \vv +
8 \pa \rr \vv
\ \ .
\end{eqnarray}
The relations
(\ref{ward2}) generalize the Ward identity (\ref{ward1}) which
is recovered for
${\cal V} = 0 = G$. They represent the {\em Ward
identities associated to the
$N=2$
super $W_3$-algebra} and they are manifestly covariant since
all differential operators occuring on their r.h.s.
are superconformally covariant when acting on the
given fields.

As a matter of fact, the Ward identities (\ref{ward2})
are equivalent to the {\em Poisson bracket algebra}
(commutation relations)
\begin{eqnarray}
\left[ \rr_2, \rr_1 \right] &=& \left( {\cal L}_2^{sym} \right) _1 \delta ^{(3)}
 (\uz _2,\uz _1)
\nonumber \\
\left[ \rr_2 ,\vv_1 \right] &=& \left( 2 \vv \pa
-(\dab \vv )D -(D\vv )\dab +(\pa \vv ) \right) _1
\delta ^{(3)} (\uz _2,\uz _1)
\label{com}
\\
\left[ \vv_2 ,\vv_1 \right] &=& -{1 \over 16}
\left({\cal L}_4^{sym} + 4 {\cal M}^{(4)sym}_{{\cal V}}
\right) _1 \delta ^{(3)} (\uz _2,\uz _1)
\ \ ,
\nonumber
\end{eqnarray}
for which we used the notation
$\rr _i \equiv \rr (\uz _i )$, 
$\vv _i \equiv \vv (\uz _i )$ and 
the Dirac distribution $\delta ^{(3)}$ defined by eq.(\ref{dirac3}).
By spelling out all expressions on the r.h.s. of eqs.(\ref{com}),
one finds that this result exactly coincides with the one found in
reference \cite{bel} by applying quite different methods
(see also \cite{pop, y} as well as \cite{dg}
 for a general formula).
The introduction of covariant operators not only allows
us to cast the results in a compact form, it also helps to exhibit
the conformal symmetry which plays a crucial role in the
formulation of integrable models. In fact, the covariant
operators occuring in eqs.(\ref{ward2}, \ref{com})
yield the second Hamiltonian structure for the $N=2$ super
Boussinesq equation \cite{bel, y, opew3}.
We note that
the algebra (\ref{com}) can be rewritten in terms of
$N=1$ superfields \cite{y} in which case all $N=2$ covariant
operators reduce to $N=1$ covariant ones 
(see \cite{gt1} for the latter operators
 and the corresponding algebra).

By integrating the Ward identities (\ref{ward2}),
one can deduce
the operator product expansions (OPE's) of
the spin $1$ and spin $2$
supercurrents $\rr$ and $\vv$. To do so, one proceeds along the
lines of reference \cite{gt2} and defines $\rr$ and ${\cal V}$ as
derivatives of a generating functional
$Z_c = Z_c \left[ H , G \right] $,
\begin{equation}
\rr = -\, \left. \frac{\delta Z_c  }{\delta H}  \right \vert _{H=0,G=0}
\qquad , \qquad
\vv
 = -  \, \left. \frac{\delta Z_c }{\delta G}  \right \vert _{H=0,G=0}
\ \ .
\end{equation}
By using
eqs.({\ref{variables}) and (\ref{dirac4}) of appendix B, one obtains
the (singular parts of the)
{\em OPE's of the
classical}  $N=2$ {\em super} $W_3${\em -algebra} \cite{opew3}:
in terms of the
notation $\rr _i \equiv \rr (\uz _i )$ and
$\vv _i \equiv \vv  (\uz _i )$, we have
 \begin{eqnarray}
\label{ope}
\rr _2 \rr _1 & \! = \! &
\frac{2}{z_{12}^2} +
\left[ \frac{\tb _{12} \th _{12}}{z_{12}^2} +
\frac{\tb _{12}}{z_{12}} \dab -
\frac{\th _{12}}{z_{12}} D-
\frac{\tb _{12} \th _{12}}{z_{12}} \pa \right] \rr _1
\nonumber \\
\rr _2 \vv  _1  & \! = \! & \left[
2 \frac{\tb _{12} \th _{12}}{z_{12}^2} +
\frac{\tb _{12}}{z_{12}} \dab -
\frac{\th _{12}}{z_{12}} D -
\frac{\tb _{12} \th _{12}}{z_{12}} \pa
\right] \vv _1 \\
\nonumber -16 \vv _2 \vv _1  & \! = \! &
\frac{12}{z_{12}^4} +
12\frac{\tb _{12} \th _{12} \rr _1}{z_{12}^4} +
12\frac{
(\tb _{12} \dab -\th _{12}D-\tb _{12} \th _{12} \pa) \rr _1}{z_{12}^3}\\
\nonumber & \! \! &
+\frac{2(20 \vv _1 +2[ D, \dab ] \rr_1 -\rr _1 ^2)-
(\tb _{12} \dab +\th _{12} D)(20 \vv _1 +8[ D, \dab ] \rr_1 -\rr _1 ^2)}
{z_{12}^2}\\ \nonumber & \! \! &
+ \frac{\tb _{12}
\th _{12} (6[ D, \dab ] \vv _1 +6\pa ^2  \rr_1 +U_1)}{z_{12}^2}
-\frac{\pa (20 \vv _1 +2[ D, \dab ] \rr_1 -\rr _1 ^2)}{z_{12}}\\
\nonumber & \! \!  &+\frac{
\tb _{12}(12\pa \dab \vv _1 +3\pa ^2 \dab \rr_1 +\bar{\psi} _1 )+
\th _{12}(12\pa D\vv _1 -3\pa ^2 D\rr_1 +\psi _1 )}{z_{12}}\\
\nonumber & \! \! &
- { 1 \over 4}
\tb _{12} \th _{12}\frac{16 \pa [ D , \dab ]  \vv _1 +
8 \pa ^3 \rr_1 +\pa [ D, \dab ]\rr _1 ^2 +
2(\pa  {\cal U} _1 -\dab  \psi _1 + D \bar{\psi} _1 )}{z_{12}}
,
\end{eqnarray}
which expressions
involve the composite supercurrents $\rr ^2$ and
${\cal U}, \bar{\psi}, \psi$ defined by
\begin{eqnarray}
{\cal U} & \! = \! &
28 \rr \vv +4 \rr [ D, \dab ] \rr -15D\rr \dab \rr -2\rr ^3
\nonumber \\
\bar{\psi} & \! = \! &
-4 \rr \dab \vv +36 \dab \rr \vv -\rr \pa \dab \rr
-\frac{3}{2} \pa \rr \dab \rr +\frac{9}{2}  [ D, \dab ] \rr \dab \rr
 -2 \rr ^2 \dab \rr  \nonumber \\
\psi & \! = \! & 4 \rr D \vv -36 D\rr \vv -\rr \pa D \rr
- \frac{3}{2} \pa \rr D\rr -\frac{9}{2} [ D, \dab ] \rr D \rr +2\rr ^2 D \rr \ \  .
\nonumber
\end{eqnarray}

In conclusion, we note that one can
explicitly derive the Poisson brackets (\ref{com})
from the OPE's (\ref{ope}) by applying the supersymmetric version of
Cauchy's theorem (appendix B) \cite{gt2}.

\section{Concluding remarks}

Interestingly enough, the zero curvature conditions provide a link
between two different classes of $N=2$ superconformally covariant
operators: by starting from the matrix representation of
sandwich operators, one obtains Ward identities which involve 
symmetric covariant operators.

Concerning the matrix representation of covariant operators 
which are not of  sandwich type, 
we remark the following:
if one disregards the covariance properties and considers
${\cal L} ^{sym} _n$ as sandwiched 
between $\dab$ and $D$, it
has the form (\ref{mgso}) and
equation (\ref{hor}) with properly chosen
coefficients $a ^{(n)} _i$ then 
provides a matrix representation.
However, this result is neither satisfactory nor complete, since it
does not properly take into account conformal symmetry and since
the scalar operator
${\cal L}^{sym}_n$ can only be recovered from this matrix representation
up to contributions ${\cal L}_a$ satisfying
$\dab {\cal L}_a D=0$.

The expression ${\cal M}^{(n)}_{{\cal W}} C$ may be viewed
as the result of a bilinear and covariant map $J$, i.e.
${\cal M}^{(n)}_{{\cal W}}C = J( {\cal W} , C)$ which is known
as the super Gordan transvectant \cite{cco}-\cite{gt1}. Such bilinear as well as
trilinear covariant operators \cite{gt1} can be constructed 
along the lines of section 3.5.
These operators  occur in particular in the Poisson brackets of super
$W$-algebras \cite{gt1}.

There is a remarkable relationship between the matrix 
representation of conformally covariant operators and 
the general formula determining the singular vectors 
of the Virasoro algebra - see the second of references  
\cite{dfiz}. This relationship generalizes to the 
$N=1$ supersymmetric theory \cite{gt1} and our results 
concerning $N=2$ covariant operators should also allow
to draw close parallels to the recent study of singular 
vectors of the $N=2$ superconformal algebra \cite{semi}.

Apart from the Gelfand-Dickey derivation of (super) 
$W$-algebras and the zero curvature construction 
that we discussed here, various other approaches to these algebras 
have been considered in the literature - see for instance
\cite{zuc, gri}. 
The tools developped in the present work should
provide the appropriate ingredients 
for generalizing these constructions
to the $N=2$ theory.

\vskip 2.4truecm
\noindent {\bf {\Large Acknowledgments}}
\vspace{3mm}

We  wish to thank Fran\c cois Delduc for informing us
about their results prior to publication, for numerous stimulating
discussions and precious advice.
F.G. acknowledges financial support by the Alexander von  
Humboldt-Foundation and 	
expresses his gratitude to Stefan Theisen for the hospitality 
extended to him in Munich at the final stage of this work.

\newpage

\appendix

\setcounter{equation}{0}
\renewcommand{\theequation}{A.\arabic{equation}}

\section*{ Appendix A :  Non-standard matrix realization of
$sl(n+1 |n)$}

$sl(n+1 |n)$ is the graded Lie algebra of
$(2n+1) \times (2n+1)$ matrices with
vanishing supertrace.
It admits a purely {\em fermionic} system of simple roots.

We consider a non-standard matrix realization of
$sl(n+1 |n)$ which is referred to as the
{\em diagonal representation} \cite{gt1, gt2}.
It consists of assigning a ${\bf Z}_2$-grading $i+j \; ({\rm mod} \, 2)$
to a matrix element $M_{ij}$,
defining the supertrace by the alternating sum
\[
{\rm str} \,  M = \sum _{i=1} ^{2n+1} (-)^{i+1} M_{ii}
\]
and the graded commutator by
\[
[M, N\}_{ik} = \sum_{j=1}^{2n+1} \left( M_{ij} N_{jk} -
(-1)^{(i+j)(j+k)} N_{ij} M_{jk} \right)
\ \ .
\]

In the Serre-Chevalley basis, the basic commutation relations
of $sl(n+1 |n)$ read
\begin{eqnarray}
\nonumber
\left[ h_i , h_j \right] &=&0
\qquad \qquad , \qquad
\left[ h_i , e_j \right] \ =\  +a_{ij} \ e_j
\\
\{ e_i , f_j \} &=& \delta _{ij} \ h_j
\qquad , \qquad
\left[ h_i , f_j \right] \ =\  -a_{ij} \ f_j
\qquad {\rm for}
\ \, i,j \in \{ 1,..,2n \}
\ \ .
\nonumber
\end{eqnarray}
Here, the $h_i$ belong to the Cartan subalgebra, the
$e_i$ denote the fermionic simple roots,
$f_i$ the associated negative roots
and the $a_{ij}$ are
the elements of the Cartan matrix:
\begin{equation}
a_{ij}=\delta _{i+1,j} - \delta_{i,j+1}
\ \ .
\end{equation}
In the diagonal representation,
the
Serre-Chevalley generators can be represented by the matrices
\begin{eqnarray}
\nonumber
h_i &=& E_{i,i} + E_{i+1,i+1}\\
e_i &=& E_{i,i+1}\\
\nonumber
f_i &=& E_{i+1,i} \ \ ,
\end{eqnarray}
where $E_{ij}$ denotes the  $(2n+1) \times (2n+1)$ matrix
with entries $(E_{ij})_{kl} = \delta _{ik} \delta _{jl} $.

The {\em superprincipal embedding of} $sl(2 | 1)$ {\em in} $sl(n+1 | n)$ is
denoted by
$sl(2 | 1) _{\pal} \subset sl(n+1 | n)$ and defined as follows
\cite{dm}:
\begin{eqnarray}
H &=& \sum_{i=1}^n (n-i+1)\, h_{2i-1}
\quad , \quad
\bar{H}\ = \ -\sum_{i=1}^n i \, h_{2i}
\nonumber
\\
\label{ppal}
\mu _+ & = & \sum_{i=1}^n e_{2i-1}
\qquad \quad \qquad \ \ , \quad
\bar{\mu} _+ \ = \ - \sum_{i=1}^n e_{2i}
\\
\mu_- & = & \sum_{i=1}^n (n-i+1) \,f_{2i-1}
\quad , \quad
\bar{\mu}_- \ = \ \ \sum_{i=1}^n i \,
 f_{2i} \ \ .
\nonumber
\end{eqnarray}
Here,
$H, \, \bar{H}$ represent the Cartan generators of $sl(2 | 1)$, $\mu _+ ,
\, \bar{\mu} _+ $
the fermionic
simple roots and
$\mu _- , \, \bar{\mu}_- $ the associated negative
roots. The non-vanishing commutators of these generators read
\begin{eqnarray}
\left[H , \bar{\mu} _{\pm} \right]& =& \pm \bar{\mu}_{\pm}
\quad \ \ \ \ , \ \quad \ \
\left[\bar{H} , \mu _{\pm} \right]\ = \ \pm \mu _{\pm}
\nonumber
\\
\{ \mu _+ , \mu _- \} &=& \ H
\qquad \ \ \ , \quad \ \
\{ \bar{\mu} _+ , \bar{\mu} _- \} \ = \ \ \bar{H}
 \\
\{ \mu _+ , \bar{\mu} _+ \} & \equiv & - E_{++}
\quad \; \ , \quad \  \
\{ \mu _- , \bar{\mu} _- \}
\ \equiv \ - E_{--}
 \ \ ,
\nonumber
\end{eqnarray}
where $E_{++}$ and $E_{--}$ correspond to
the bosonic roots of $sl(2 | 1)$ .

The Lie superalgebra
$sl(2|1)_{\pal}$ contains the
superprincipal embedding of $osp(1 | 2)$ into
$sl(n+1 | n)$, i.e.
$osp(1 | 2)_{\pal} \subset sl(2|1)_{\pal} \subset sl(n+1|n)$: this
embedding is defined by the combinations
\begin{eqnarray}
J_0  =H+\bar{H} \quad ,\quad
\kappa _+ & \! = & \! \mu _+ + \bar{\mu} _+ \quad ,
\quad \kappa _- = \mu _- + \bar{\mu} _-
\nonumber
\\
J_+ & \! = & \! E_{++}
\qquad \ \ , \quad \, J_- = -E_{--}
\ \ ,
\end{eqnarray}
which satisfy the commutation relations
\begin{equation}
\left[ J_0 , \kappa _{\pm} \right] = \pm  \kappa _{\pm} \ \ ,
\ \ \{ \kappa _+ , \kappa _- \} = J_0 \ \ ,
\ \ \{ \kappa _{\pm} , \kappa _{\pm} \} = \pm 2 J_{\pm}  \ \ .
\end{equation}
Here, $J_0$ is the Cartan generator and
$\kappa _+$ the fermionic
simple root.
The highest weight generators $M_k$ of
$osp(1 | 2)_{\pal}$
are defined by the relations
\begin{equation}
\left[ J_0 , M_k \right] = k \ M_k \qquad ,\qquad  [ \kappa _+ , M_k \}
= 0 \qquad \ (\, k  = 1,2,...,2n \, )
\ \ ,
\end{equation}
which are solved by
\begin{equation}
M_k = M_1 ^k \qquad \ \ {\rm with} \ \ \ M_1 = \mu _+ - \bar{\mu} _+ \ .
\end{equation}

\section*{Appendix B: Some distributional relations}

\setcounter{equation}{0}
\renewcommand{\theequation}{B.\arabic{equation}}

In this appendix,
we gather some distributional relations which hold
in $d=2, \, N=2$ superspace
and we indicate 
the correspondence rules between supercurrent
OPE's and commutation relations.

Let
$(\uz_i,\bar{\uz}_i) \equiv (z_i,\th_i,\tb_i \, ;\zb_i,\thm_i,\tbm_i)$
with $i=1,2$ denote the
local coordinates of two points on a $N=2$ SRS.
We then define the relative coordinates by 
\begin{equation}
\label{variables}
\th _{12}  =  \th _1 - \th _2  \quad ,\quad  \tb _{12} = \tb _1 - \tb _2
\quad , \quad
z_{12} = z_1 -z_2 -\frac{1}{2} (\th _1 \tb _2 +\tb _1 \th _2 )
 \ .
\end{equation}

For a superanalytic
superfield $F$ (i.e. a field
depending only on $\uz$),
one has the following Cauchy formulae
\cite{West, pop} ($n \in {\bf N}$):
\begin{equation}
\label{Cauchy}
\ds \int d^3 \uz_1
\left(
\tb _{12} \th _{12} , \th_{12} , \tb_{12} , 1 \right)
{F(\uz_1) \over
z_{12}^{n+1} }
= \left( 1 , -\dab , D , {1\over 2 } [D, \dab ] \right)
{\pa ^n F(\uz_2) \over n !}
\ \ .
\end{equation}
Here, $d^3 \uz$ denotes the measure $(2\pi i)^{-1} dzd\th d\tb$
with respect to which the
Dirac distribution takes the form
\begin{equation}
\label{dirac3}
\delta ^{(3)} (\uz _1,\uz _2) \equiv 
\delta  (z_1 -z_2 )  (\tb _1 -\tb _2)(\th _1 -\th_2) \ .
\end{equation}

The integration of Ward identities is performed by means of
the Dirac distribution defined with respect to the measure
 $d^4 \uz = ( 2\pi i)^{-1}  d\zb dzd\th d\tb$:
\begin{equation}
\label{dirac4}
\delta ^{(4)} (\uz _1,\uz _2) \equiv
\delta (z_1 -z_2) \delta (\zb_1-\zb_2 )  (\tb _1 -\tb _2)(\th _1 -\th_2)
=\pa _{\zb _1 }
\frac{1}{z_{12}} \, \tb _{12} \th _{12} \ .
\end{equation}

A general OPE of two supercurrents $A(\uz _2)$
and $B(\uz _1)$ can be written in the form
\begin{equation}
\label{opeAB}
A(\uz _2)B(\uz _1) =\sum _{n \in {\bf N}}
\frac{a_n(\uz_1)+\tb_{12} \alpha_n(\uz_1)+\th_{12} \beta_n(\uz_1)+
\tb _{12} \th_{12} b_n(\uz_1)}{z_{21} ^{n+1}}
\ +\ {\rm regular} \ {\rm terms}
\end{equation}
and by virtue of the relations (\ref{Cauchy}), this OPE
is equivalent to the commutation relation
\begin{equation}
\label{comAB}
\left[ A(\uz _2),B(\uz _1) \right] =\sum _{n \in {\bf N}}
\frac{1}{n !} \left. \left(
\frac{a_n(\uz)}{2} [ D, \dab ]+\alpha_n(\uz) D -
\beta_n(\uz) \dab +b_n (\uz) \right) \right \vert _{\uz =\uz _1}
\pa_1 ^n \delta ^{(3)} (\uz _2,\uz _1)\
 .
\end{equation}

\end{document}